\RequirePackage[hyphens]{url} 
\documentclass[a4paper,fleqn,usenatbib]{mnras}

\usepackage{newtxtext,newtxmath}

\usepackage[T1]{fontenc}
\usepackage{ae,aecompl}

\usepackage{graphicx}	
\usepackage{amsmath}	
\usepackage{longtable}

\newcommand{\txn}{\textnormal}
\newcommand{\us}{\,} 

\newcommand{\code}{\nolinkurl} 

\newcommand{\eagle}{EAGLE}

\newcommand{\Ang}{\ensuremath{\us \txn{\AA}}}

\newcommand{\cMpc}{\ensuremath{\us \txn{cMpc}}} 
\newcommand{\ckpc}{\ensuremath{\us \txn{ckpc}}} 
\newcommand{\pkpc}{\ensuremath{\us \txn{pkpc}}} 
\newcommand{\pcmsq}{\ensuremath{\us \txn{cm}^{-2}}}
\newcommand{\kmps}{\ensuremath{\us \txn{km}\, \txn{s}^{-1}}}
\newcommand{\pcc}{\ensuremath{\us \txn{cm}^{-3}}}
\newcommand{\Msun}{\ensuremath{\us \txn{M}_{\sun}}}
\newcommand{\Lstar}{\ensuremath{\us \txn{L}_{*}}}

\newcommand{\Mvir}{\ensuremath{\us \txn{M}_{\txn{200c}}}}
\newcommand{\Rvir}{\ensuremath{\us \txn{R}_{\txn{200c}}}}
\newcommand{\Tvir}{\ensuremath{\us \txn{T}_{\txn{200c}}}}

\newcommand{\dex}{\ensuremath{\us \txn{dex}}}
\newcommand{\K}{\ensuremath{\us \txn{K}}}

\title[X-ray line emission in the EAGLE CGM]{The warm-hot circumgalactic medium around EAGLE-simulation galaxies and its detection prospects with X-ray line emission}

\author[N.\ A.\ Wijers \& J. Schaye]{
Nastasha A.\ Wijers,$^{1,2}$\thanks{E-mail: nastasha.wijers@northwestern.edu}
Joop Schaye$^{1}$ 
\\
$^{1}$Leiden Observatory, Leiden University,
PO Box 9513, NL-2300 RA Leiden, The Netherlands\\
$^{2}$CIERA and Department of Physics and Astronomy, Northwestern University, 1800 Sherman Ave, Evanston, IL 60201, USA\\
}

\date{Accepted XXX. Received YYY; in original form ZZZ}

\pubyear{2022}

\begin{document}
\label{firstpage}
\pagerange{\pageref{firstpage}--\pageref{lastpage}}
\maketitle

\begin{abstract}
We estimate the detectability of X-ray metal-line emission from the circumgalactic medium (CGM) of galaxies over a large halo mass range ($\Mvir=10^{11.5}$--$10^{14.5}\Msun$) using the EAGLE simulations. With the XRISM Resolve instrument, a few bright (K-$\alpha$ or Fe L-shell) lines from $\Mvir \gtrsim 10^{13}  \Msun$ haloes should be detectable. Using the Athena X-IFU or the Lynx Main Array, emission lines (especially from \ion{O}{viii} and \ion{O}{vii}) from the inner CGM of $\Mvir\gtrsim10^{12.5}\Msun$ haloes become detectable, and intragroup and intracluster gas will be detectable out to the virial radius. With the Lynx Ultra-high Resolution Array, the inner CGM of haloes hosting $\Lstar$ galaxies is accessible. These estimates do assume long exposure times ($\sim 1$~Ms) and large spatial bins ($\sim1$--$10\us\mathrm{arcmin}^{2}$).   
This emission is dominated by collisionally ionized (CI) gas, and tends to come from halo centres. The emission is biased towards temperatures close to the maximum emissivity temperature for CI gas ($\mathrm{T}_\mathrm{peak}$), and high densities and metallicities.
However, for the K-$\alpha$ lines, emission can come from hotter gas in haloes where the virialized, volume-filling gas is hotter than $\mathrm{T}_\mathrm{peak}$. 
Trends of emission with halo mass can largely be explained by differences in virial temperature. 
Differences in the mass trends of K-$\alpha$, He-$\alpha$-like, and Fe~L-shell lines mirror differences in their emissivities as a function of temperature.
We conclude that upcoming X-ray missions will open up a new window onto the hot CGM. 
\end{abstract}




\section{Introduction}

In order to understand the formation, evolution, and quenching of galaxies, we must understand the gas that surrounds them: the circumgalactic medium (CGM). 
This is because, first, accretion of gas from the CGM fuels star formation in star-forming galaxies. Without the additional gas supply, star-forming galaxies would deplete their (galactic) gas reservoirs on timescales too short to be consistent with their star formation histories \citep[e.g., the CGM review by][]{tumlinson_peeples_werk_2017_cgmreview}. 
In numerical simulations, cutting off this fuel supply is one way to quench star formation in galaxies \citep[e.g.,][]{oppenheimer_davies_etal_2020, zinger_pillepich_etal_2020}.
Secondly, galaxies inject mass and metals back into the CGM. Outflows from galaxies into the CGM have been observed \citep[e.g., the review by][]{rupke_2018}, and are required to explain the presence of metals in the diffuse intergalactic medium (IGM) \citep[e.g.,][]{aguirre_hernquist_etal_2001, booth_schaye_etal_2012}.
In numerical  simulations, these outflows (driven by e.g., supernovae and AGN) are required to reproduce the galaxy stellar mass function \citep[e.g.,][]{cole_1991, hopkins_keres_etal_2013, eagle_calibration, pillepich_springel_etal_2018}. Therefore, observing the CGM can teach us about the processes that drive, regulate, and quench star formation.

Around isolated galaxies at low redshift, especially at $\sim \Lstar$, much of what we know about the CGM comes from studies of UV absorption lines, often using the \emph{Hubble Space Telescope's} \emph{Cosmic Origins Spectrograph} (HST-COS) \citep[{e.g.,}][]{tumlinson_thom_etal_2011, johnson_chen_mulchaey_2015, johnson_chen_etal_2016}. These lines mainly trace cool to warm ($\sim 10^{4}$--$10^{5.5}$~K) gas. 

A few ions producing UV lines trace warmer gas: \ion{O}{vi} is sensitive to $\sim 10^{5.5}$~K gas if it is collisionally ionized. However, simulations 
\citep[{e.g.,}][]{tepper-garcia_richter_etal_2011, rahmati_etal_2016, oppenheimer_etal_2016, oppenheimer_2018_fossilAGN_cos, roca-fabrega_dekel_etal_2018, wijers_schaye_oppenheimer_2020}
and observations 
\citep[{e.g.,}][]{carswell_schaye_kim_2002, trip_sembach_etal_2008, werk_proschaska_etal_2014, werk_prochaska_etal_2016} 
alike suggest both collisionally ionized and photo-ionized \ion{O}{vi} is present in the CGM, and causes measurable absorption lines, complicating the interpretation of observations. The \ion{Ne}{VIII} ion produces a doublet in the extreme UV (EUV) range, but at redshift $\gtrsim 0.5$, it redshifts into the far UV (FUV) energy band and can be observed. This ion has been used to study the hotter CGM ($\sim 10^{5}$--$10^{6}$~K), by e.g., \citet{burchett_tripp_etal_2018} (observationally) and \citet{tepper-garcia_richter_etal_2013} (in a cosmological simulation). 

However, much of the gas in the CGM of low-redshift galaxies is expected to be at higher temperatures \citep[$> 10^{5.5}$--$10^{6}$~K, e.g.,][]{wijers_schaye_oppenheimer_2020}. We expect this hot gas to be present around $\sim \Lstar$ and more massive galaxies ($\log_{10} \Mvir \Msun^{-1} \gtrsim 11.5$--$12.0$), where a volume-filling, virialized gas phase has formed  
\citep[e.g.,][]{dekel_birnboim_2006, keres_katz_etal_2009, van-de-voort_schaye_etal_2011, correa_schaye_etal_2018}.
Aside from some EUV lines, this gas produces most of its emission \citep[e.g.,][]{bertone_aguirre_schaye_2013} and absorption \citep[e.g.,][]{perna_loeb_1998, hellsten_gnedin_miralda-escude_1998} lines in X~rays. 

Therefore, X-ray emission and absorption are one way we might detect the hot phase of the CGM. X-ray emission and absorption lines in this hot gas come from metals, as hydrogen and helium are fully ionized at such high temperatures. Besides these lines, the warm-hot gas produces X-ray continuum emission. \citet{bertone_aguirre_schaye_2013} predicted that most X-ray emission from diffuse gas throughout the universe is in the form of continuum emission. However, around isolated ellipticals and in groups, X-ray emission is typically line dominated \citep[e.g., the review by][]{werner_mernier_2020}.

Other observables of the warm/hot CGM include the dispersion measures of fast radio bursts (FRBs). These are sensitive to the free electron density along the line of sight, although they only measure the total electron column density \citep[e.g.,][]{prochaska_zheng_2019}, meaning they are equally sensitive to photo-ionized, cool gas and warm-hot, collisionally ionized gas. They can be used to constrain the ionized gas content of haloes, but this requires a sufficiently large sample of FRBs with known redshifts \citep{ravi_2019}.

The Sunyaev-Zel'dovich (SZ) effect is also sensitive to free electrons: the thermal SZ effect probes the electron pressure along the line of sight, and the kinetic SZ effect measures the electron bulk velocity. These effects have primarily been used to study clusters \citep[e.g., the review by][]{mroczkowski_nagai_etal_2018}. However, tSZ signals from massive filaments have also been detected by stacking pairs of massive galaxies \citep[e.g.,][]{de-graaff_cai_etal_2017, tanimura_hinshaw_etal_2019}, and studies of lower mass haloes have been done by fitting models, using the known positions of galaxy groups \citep[e.g.,][]{lim_mo_etal_2018, lim_mo_etal_2020}. 
Both kinetic and thermal SZ signals from low mass systems are difficult to study with current instruments due to their large beam size \citep[spatial resolution; e.g., ][]{mroczkowski_nagai_etal_2018}.  

Around massive galaxies and in groups and clusters, X-ray emission from the CGM, intra-group medium (IGrM) and intra-cluster medium (ICM) has been detected \citep[e.g., the review by][]{werner_mernier_2020}. For isolated galaxies, these detections are mostly limited to massive (elliptical) galaxies. For lower-mass, spiral galaxies, studies have typically found upper limits or emission only in or close to galaxies \citep[e.g.,][]{bogdan_vogelsberger_etal_2015}. However, \citet{das_mathur_gupta_2020} found emission further from the galaxy, and measured a temperature profile out to $\approx 200$~kpc.

Another exception is the Milky Way. The halo of our own Galaxy has been studied using X-ray line emission, often in combination with X-ray absorption lines \citep[e.g.,][]{bregman_lloyd-davies_edward_2007,  gupta_mathur_etal_2014, miller_bregman_2015, das_mathur_etal_2019}. Other studies focussed on absorption lines \citep[e.g.,][]{kuntz_snowden_2000, hodges-kluck_miller_bregman_2016, gatuzz_churazov_etal_2018}. These measurements have been used to constrain e.g., the hot phase temperature \citep[e.g.,][]{kuntz_snowden_2000, das_mathur_etal_2019} and halo rotation \citep{hodges-kluck_miller_bregman_2016}. It is not certain how extended the gas causing the absorption and emission is \citep[e.g.,][]{bregman_lloyd-davies_edward_2007, gatuzz_churazov_etal_2018}, though \citet{miller_bregman_2015} placed some constraints on the density profile and metallicity using a combination of \ion{O}{vii} and \ion{O}{viii} absorption and emission lines.

X-ray emission has been very useful in the study of the ICM. From spectra, the temperature, electron density, and element abundances (using the ratio of emission lines and continuum) of the X-ray emitting phase have been measured \citep[e.g., the review by][]{werner_mernier_2020}. Turbulence has also been measured, using resonant scattering emission lines \citep[e.g., the review by][]{werner_mernier_2020}, spatially resolved emission line profiles \citep[e.g.,][]{hitomi-collaboration_ahoranian_etal_2018}, and surface brightness fluctuations \citep[e.g.,][]{zhuravleva_churazov_etal_2014}.

Extending such studies toward lower halo masses would be very valuable. The mass of the CGM around e.g., $\Lstar$ galaxies is very uncertain, especially the mass of the warm/hot gas \citep[e.g.,][fig.~11]{werk_proschaska_etal_2014}. Theoretical predictions also differ hugely: for example, the EAGLE and IllustrisTNG simulations predict very different CGM gas masses in $\lesssim \Lstar$ haloes \citep{davies_crain_etal_2019_tngcomp}, even though both produce broadly realistic galaxy populations \citep[e.g.,][]{eagle_paper, pillepich_springel_etal_2018}. \citet{oppenheimer_bogdan_etal_2020} have found that broad-band X-ray emission from the EAGLE and IllustrisTNG $\Lstar$ inner CGM should be observable with eRosita, and that this instrument should be able to distinguish between the two models. 

Many predictions of X-ray emission from hot haloes from numerical simulations have focussed on groups and clusters \citep[e.g.,][]{barnes_kay_etal_2017, truong_rasia_etal_2018, cucchetti_pointecouteau_etal_2018, mernier_cucchetti_etal_2020}, where much of the data is currently available and high photon counts will allow detailed information to be extracted. For {\eagle}, \citet{eagle_paper} have studied X-ray emission from groups and clusters, and \citet{davies_crain_etal_2019} considered broad-band soft X-ray emission over a large range of halo masses and found it to be a good diagnostic of the CGM gas mass at fixed halo mass.  \citet{bogdan_vogelsberger_etal_2015} made predictions of broad-band X-ray emission from the Illustris simulation \citep{vogelsberger_genel_etal_2014_methods} 
and compared them to data, finding them to be broadly consistent. \citet{zhang_cui_etal_2020} made predictions of X-ray emission for HUBS \citep{cui_chen_etal_2020, cui_bregman_etal_2020} observations across a large range of halo masses using IllustrisTNG \citep{pillepich_springel_etal_2018}, 
and \citet{truong_pillepich_etal_2020} related hot gas properties close to the central galaxy to the central galaxy properties in IllustrisTNG. 
\citet{van-de-voort_schaye_2013} made predictions for X-ray line emission specifically, using the OWLS simulations \citep{schaye_dalla-vecchia_etal_2010}. We update these predictions and extend them to a larger set of emission lines using {\eagle}. 
   
In this paper, we will study low-redshift ($z=0.1$) X-ray emission lines as predicted using the {\eagle} simulations. We describe the simulations and how we use them to predict line emission in \S\ref{sec:methods}. We select a number of the stronger emission lines we expect to find (\S\ref{sec:lines}), and compare them to estimated sensitivity limits of various planned and proposed X-ray telescopes (\S\ref{sec:results}). We describe how we estimate those sensitivity limits in \S\ref{sec:det}. We also investigate the gas responsible for the emission and how it compares to typical CGM gas in \S\ref{sec:results}. In \S\ref{sec:discussion}, we discuss our results, and we summarise them in \S\ref{sec:conclusions}. 
For a similar study of X-ray and highly ionized UV \emph{absorption} lines (\ion{O}{vi}--\ion{}{viii}, \ion{Ne}{viii}, \ion{Ne}{ix}, and \ion{Fe}{xvii}) in the CGM of {\eagle} galaxies, see  \citet{wijers_schaye_oppenheimer_2020}.  

Note that we will often use `CGM' or `halo' as a catch-all term for what is typically called the CGM (gas around isolated galaxies), as well the IGrM and the ICM in the few clusters in the $100^{3} \cMpc^{3}$ {\eagle} volume. We describe distances as comoving (e.g., `cMpc') or proper/physical (e.g., `pkpc'), except for centimetres, which are always physical. We use a Lambda cold dark matter cosmogony with the \citet{planck_2013} cosmological parameters: $(\Omega_m,\Omega_\Lambda,\Omega_b, h, \sigma_8, n_s, Y) = (0.307, 0.693, 0.04825, 0.6777, 0.8288, 0.9611, 0.248)$. These are the same values as were used in the {\eagle} simulations.

\section{Methods}
\label{sec:methods}

In this section, we will discuss the cosmological simulations we use to make our predictions, how we predict surface brightnesses from them, and the galaxy and halo information we use.

\subsection{{\eagle}}
\label{sec:eagle}

We study line emission using the {\eagle}  \citep[`Evolution and Assembly of GaLaxies and their Environments';][]{eagle_paper, eagle_calibration, mcalpine_helly_etal_2016} cosmological, hydrodynamical simulations. Specifically, we use the \code{Ref-L0100N1504} $100^{3} \cMpc^{3}$ volume, with an initial gas mass resolution of  $1.81 \times 10^6 \Msun$ and a Plummer-equivalent gravitational softening length of $0.70 \pkpc$ (at low redshift, like we study here). {\eagle} uses a modified version of  Gadget3  \citep{springel_2005}; gravitational forces are calculated using the Tree-PM scheme, and hydrodynamical forces are calculated using a pressure-entropy formulation of SPH known as Anarchy (\citeauthor{eagle_paper} \citeyear{eagle_paper}, appendix~A; \citeauthor{anarchy_effect} \citeyear{anarchy_effect}). 

Besides gravity and hydrodynamics, {\eagle} also models the effects of processes that occur on scales below its resolution: so-called subgrid physics. Radiative cooling and heating is modelled as described by \citet{wiersma_schaye_smith_2009}, including the effects of 9 metal abundances tracked in {\eagle}. Because the resolution is too low to model the multi-phase ISM, molecular cooling and heating channels are not included, and artificial fragmentation of the interstellar medium (ISM) is prevented by setting a pressure floor in dense gas that ensures the Jeans mass remains marginally resolved \citep{eagle_paper}. This means the temperature of star-forming gas is set by the pressure floor equation of state, and is generally not typical of what we expect for the ISM. 

Stars form stochastically in sufficiently dense gas, with a threshold that depends on the gas metallicity \citep{schaye_2004}. The star-formation rate depends on pressure in a way that, by design, reproduces the Kennicutt-Schmidt relation \citep{schaye_dalla-vecchia_2008}. Feedback from these stars is modelled as well. Core-collapse supernovae inject thermal energy stochastically into neighbouring gas particles \citep{dalla-vecchia_schaye_2012}. The thermal energy injection raises the gas temperature by $10^{7.5} \K$, with a probability set to match the (calibrated) supernova energy budget per unit stellar mass. Core-collapse supernovae, as well as AGB winds and type~Ia supernovae, inject mass and metals into the surrounding gas, with metal yields for 9 individual elements following \citet{wiersma_etal_2009_insim}. 

Black holes are seeded in sufficiently massive haloes that do not already contain them. They can merge and accrete gas following \citet{rosas-guevara_bower_etal_2015}. Black holes generate AGN feedback by thermal energy injection \citep{booth_schaye_2009}, like supernovae, but raise the gas temperature by $10^{8.5} \K$. 


Because the way the feedback energy couples to gas on scales resolved in {\eagle} is still uncertain, the feedback on resolved scales is calibrated to produce realistic galaxies. The supernova and black hole feedback is calibrated to match the $z=0.1$ galaxy luminosity function and stellar-mass-black-hole-mass relation, and to produce reasonable galaxy sizes \citep{eagle_calibration}.  The {\eagle} simulation data has been publicly released, as described by \citet{mcalpine_helly_etal_2016} and \citet{eagle-team_2017}.

\subsection{The emission lines}
\label{sec:lines}

\subsubsection{The default tables}

We will describe the luminosity and surface brightness of the CGM in EAGLE for a set of soft X-ray emission lines.
The basis for our selection of X-ray lines is a set of tables from \citet{bertone_schaye_etal_2010}, and the lines studied in that work. These tables were calculated using {\textsc CLOUDY} v7.02 \citep[last documented in][]{cloudy}, assuming a \citet{HM01} uniform, but redshift-dependent, UV/X-ray ionizing background. Note that this means that, when calculating the emission from a patch of gas, we ignore contributions to the incident radiation field from, e.g., nearby AGN or ICM. This is consistent with the radiative cooling used in the EAGLE simulation. 

Following Charlotte Brand (private communication, 2017), our lines were selected to have peak emissivities in dense gas between $10^{5}$ and $10^{7}$~K, as this is the warm-hot gas phase we want to investigate. The lines have energies $> 0.3$~keV, based on absorption by our own Galaxy (see e.g., the left panel of Fig.~\ref{fig:ins}).  

The lower line energy limit constrains the emissivity peaks of these lines to be at $\gtrsim 10^{6}$~K (Fig.~\ref{fig:emcurves}), while higher-energy lines such as the \ion{Si}{xiv} Lyman-$\alpha$-like (K-$\alpha$) line and the \ion{S}{xv} He-$\alpha$-like recombination line are excluded based on their temperature peaks. In addition to the selection of bright lines from \citet{bertone_schaye_etal_2010}, we also include more iron L-shell lines and the \ion{Mg}{xi} He-like resonance line.
Similarly, the \ion{Si}{ix}--\ion{}{xii} ions produce a few emission lines between $0.3$ and $0.4$~keV, the brightest of which matches the peak emissivity of the \ion{Si}{xiii} He-like resonance line. However, this is still fainter than the carbon lines in this energy range. We choose to focus instead on the He-$\alpha$-like and K-$\alpha$ transitions, and a set of relatively bright iron L-shell lines. 
Following \citet{bertone_schaye_etal_2010}, we mostly consider only the resonance line for He-$\alpha$-like triplets, except for the brightest one, \ion{O}{vii}. There, we consider the forbidden (f) and intercombination (x and y) lines as well. 
We list the lines we study 
in Tables~\ref{tab:lines} and~\ref{tab:linesPS20}.

\begin{table*}
\caption{Data for the lines from the \citet{bertone_schaye_etal_2010} tables we investigate in this work: the ion producing the line, its wavelength and energy (rest-frame), the peak line emissivity for collisional ionization equilibrium (CIE) conditions ($\mathrm{n}_{\mathrm{H}} = 10 \pcc$), using solar metal abundances (Table~\ref{tab:Zsol}), the temperature at the CIE emission peak, and the temperature range in which the emissivity is at least 0.1 times the maximum value. The peak temperatures are the maxima obtained directly from the emissivity tables, which use a 0.05~dex temperature grid. The final three columns list data for comparing these lines to other works: the electron configurations of the upper and lower levels of the transition, and names of the lines in {\sc CLOUDY}~v7.02. We substitute underscores for spaces in the names. Note that the line energies we list here are derived from the {\sc CLOUDY}~v7.02 wavelengths; they are listed for convenience.}
\label{tab:lines}
\begin{tabular}{l l l l l c l l l}
\hline
ion & $\lambda$ & E &$\max \, \Lambda \,\mathrm{n}_\mathrm{H}^{-2} \mathrm{V}^{-1}$ & $\mathrm{T}_{\mathrm{peak}}$ & $\mathrm{T}_{0.1 \times \mathrm{peak}}$ & upper level & lower level  &  name\\
 & $\textnormal{\AA}$ & keV & $\log_{10} \, \mathrm{erg} \, \mathrm{cm}^{3} \mathrm{s}^{-1}$ & $\log_{10}$~K & \ $\log_{10}$~K & & & {\sc CLOUDY} v7.02 \\
\hline
\ion{C}{v} 		& 40.27 & 0.3079 & -24.4 & 5.95 & 5.7--6.3 & 1s 2p $^1$P & 1s$^2$ $^1$S$_0$ & C\_\_5\_40.27A\\
\ion{C}{vi}		& 33.74 & 0.3675 & -24.1 & 6.15 & 5.9--6.8 & 2p & 1s & C\_\_6\_33.74A \\
\ion{N}{vi} 		& 28.79 & 0.4307 & -24.7 & 6.15 & 5.9--6.5 & 1s 2p $^1$P & 1s$^2$ $^1$S$_0$ & N\_\_6\_28.79A \\
\ion{N}{vii} 	& 24.78 & 0.5003 & -24.4 & 6.3   & 6.1--7.0 & 2p & 1s & N\_\_7\_24.78A\\
\ion{O}{vii} 	& 21.60 & 0.5740 & -23.9 & 6.3   & 6.0--6.7 & 1s 2p $^1$P & 1s$^2$ $^1$S$_0$ & O\_\_7\_21.60A\\
\ion{O}{vii} 	& 21.81 & 0.5685 & -24.4 & 6.35 & 6.0--6.7 & 1s 2p $^3$P & 1s$^2$ $^1$S$_0$ & O\_\_7\_21.81A\\
\ion{O}{vii} 	& 22.10 & 0.5610 & -23.9 & 6.35 & 6.0--6.7 & 1s 2s $^3$S & 1s$^2$ $^1$S$_0$ & O\_\_7\_22.10A\\
\ion{O}{viii} 	& 18.97 & 0.6536 & -23.6 & 6.5   & 6.2--7.2 & 2p & 1s & O\_\_8\_18.97A\\
\ion{Ne}{ix}	& 13.45 & 0.9218 & -24.4 & 6.6   & 6.3--7.0 & 1s 2p $^1$P & 1s$^2$ $^1$S$_0$ & Ne\_9\_13.45A\\
\ion{Ne}{x}	& 12.14 & 1.021   & -24.2 & 6.8   & 6.5--7.5 & 2p & 1s & Ne10\_12.14A\\
\ion{Mg}{xi}	& 9.169 & 1.352   & -24.8 & 6.8   & 6.4--7.2 & 1s 2p $^1$P & 1s$^2$ $^1$S$_0$ & Mg11\_9.169A \\
\ion{Mg}{xii}	& 8.422 & 1.472   & -24.6 & 7.0   & 6.7--7.8 & 2p & 1s & Mg12\_8.422A\\
\ion{Si}{xiii} 	& 6.648 & 1.865   & -24.8 & 7.0   & 6.6--7.4 & 1s 2p $^1$P & 1s$^2$ $^1$S$_0$ & Si13\_6.648A\\
\hline
\end{tabular}
\end{table*}

\begin{table*}
\caption{Data for the \citet{ploeckinger_schaye_2020} table lines we use in this work, analogous to the Table~\ref{tab:lines} data. For the peak emissivity, the metallicity of the gas is scaled to the solar values of {\sc CLOUDY}~v7.02 (Table~\ref{tab:Zsol}).  The peak temperatures are the maxima directly from the emissivity tables, which use a 0.1~dex temperature grid. The line identifications/transitions come from 
the {\sc Chianti} database (versions 7.0 and 10.0.1) used in {\sc CLOUDY}~v17.01 for these iron lines. Note that the line energies we list here are derived from the {\sc CLOUDY} wavelengths; they are listed for convenience. The transition attributed to the \ion{Fe}{xvii}~17.05$\Ang$ line in the {\sc Cloudy}~v7.02 data produces a line at 16.7757~$\Ang$ in the Chianti database used to match the Fe~L-shell lines in {\textsc CLOUDY}~v17.01. However, we believe the comparison between both lines at 17.05~$\Ang$ is more like-for-like (see the text for discussion).}
\label{tab:linesPS20}
\begin{tabular}{l l l l l c l l l}
\hline
ion 	& $\lambda$ 	& E 	& $\max \, \Lambda \,\mathrm{n}_\mathrm{H}^{-2} \mathrm{V}^{-1}$ 	& $\mathrm{T}_{\mathrm{peak}}$ 	& $\mathrm{T}_{0.1 \times \mathrm{peak}}$& upper level 	& lower level 	& name \\
 	& $\textnormal{\AA}$ 	& keV 	& $\log_{10} \, \mathrm{erg} \, \mathrm{cm}^{3} \mathrm{s}^{-1}$ 	&  $\log_{10}$~K 	&  $\log_{10}$~K 	&  	&  	& {\textsc CLOUDY}~v17.01 \\
\hline
Fe XVII 	&       17.0960 	& 0.7252 	& -24.1 	& 6.7 	& 6.3--7.0 	& 2s$^2$ 2p$^5$ 3s $^3$P$_2$ 	& 2s$^2$ 2p$^6$  $^1$S$_0$  & Fe17\_\_\_\_\_\_17.0960A \\
Fe XVII 	&       17.0510 	& 0.7271 	& -24.0 	& 6.7 	& 6.3--7.0 	& 2s$^2$ 2p$^5$ 3s $^1$P$_1$ 	& 2s$^2$ 2p$^6$  $^1$S$_0$  & Fe17\_\_\_\_\_\_17.0510A \\
Fe XVII 	&       16.7760 	& 0.7391 	& -24.2 	& 6.7 	& 6.4--7.0 	& 2s$^2$ 2p$^5$ 3s $^3$P$_1$ 	& 2s$^2$ 2p$^6$  $^1$S$_0$  & Fe17\_\_\_\_\_\_16.7760A \\
Fe XVII 	&       15.2620 	& 0.8124 	& -24.3 	& 6.8 	& 6.4--7.0 	& 2s$^2$ 2p$^5$ 3d $^3$D$_1$ 	& 2s$^2$ 2p$^6$  $^1$S$_0$  & Fe17\_\_\_\_\_\_15.2620A \\
Fe XVIII 	&       16.0720 	& 0.7714 	& -24.9 	& 6.8 	& 6.5--7.1 	& 2s$^2$ 2p$^4$ ($^3$P) 3s $^4$P$_{5/2}$ & 2s$^2$ 2p$^5$ $^2$P$_{3/2}$ & Fe18\_\_\_\_\_\_16.0720A \\
\hline
\end{tabular} 
\end{table*}

We looked up the transitions for the bright iron L-shell lines from {\sc CLOUDY}~7.02 in the Opacity Project database\footnote{\url{http://cdsweb.u-strasbg.fr/topbase/topbase.html}}, described as the source of the L-shell lines (included in  {\sc CLOUDY} v7.02 via \code{level2.dat}). The exception is the \ion{Fe}{xvii} $17.05 \Ang$ transition, which we could not find in that database. We found the data for that transition by comparing the wavelength (and checking the weighted oscillator strength $gf$) to the line compilation of \citet{mewe_gronenschild_1981} and the lines in the NIST database{\footnote{\url{https://physics.nist.gov/asd}, accessed 2020-09-24}} \citep{shorer_1979, gordon_hobby_peacock_1980}.

The K-$\alpha$ and He-$\alpha$-like transitions are calculated internally in {\sc CLOUDY}~v7.02. We looked up the He-like transitions in the documentation (\code{LineList_He_like.dat}). The K-$\alpha$ transitions include both doublet components (all 2p -- 1s transitions). The 2s~$^3$S -- 1s$^2$~$^1$S$_0$ transition is the He-like forbidden line, 2p~$^3$P -- 1s$^2$~$^1$S$_0$ are the He-$\alpha$-like intercombination lines (sum of the x and y lines), and 2p~$^1$P -- 1s$^2$~$^1$S$_0$ is the He-$\alpha$-like resonance line. \citet{porquet_dubau_2000} give a helpful overview of these He-like transitions in their fig.~1.




\subsubsection{The Fe L-shell lines}

For the bright iron L-shell lines we investigated, the \citet{bertone_schaye_etal_2010} tables in the previous section contain an error: at densities $\mathrm{n}_{\mathrm{H}} \lesssim 10^{-4} \pcc$ (for \ion{Fe}{xvii} $17.05 \Ang$) and $\lesssim 10^{-3}$--$10^{-2} \pcc$ (for the other L-shell lines), there is no tabulated emission at all from these lines. The tabulated quantity is $\log_{10} \mathrm{L} \, \mathrm{V}^{-1} \mathrm{n}_{\mathrm{H}}^{-2}$, where L is luminosity, V is volume, and $\mathrm{n}_{\mathrm{H}}$ is the hydrogen number density. This means that, in the collisionally ionised (CI) limit, the tabulated quantity should not depend on density. The drop in tabulated emissivity is a very sharp transition from emission in the CI limit to zero, so the drop is not a physically consistent decrease of emission to (floating point) zero at low densities. This behaviour is not the result of some physical process.

This bug also affects the  \ion{Fe}{xvii} $17.05 \Ang$ analysis of \citet{bertone_schaye_etal_2010}, as can be seen in e.g., their fig.~13. There is some slightly lower-density gas producing \ion{Fe}{xvii} $17.05 \Ang$ emission in that plot, at $z=0.25$, than we find at $z=0.1$. However, the lack of any emission at all at temperatures and densities where there is emission in the other metal lines indicates that there is a similar issue at this redshift.

In order to make predictions for these lines, we therefore use a different set of tables: those of \citet{ploeckinger_schaye_2020}. We use their default tables, which include the effects of dust, self-shielding, cosmic rays, and local stellar radiation. However, these dense gas/ISM processes and effects are negligible for the X-ray lines we study here. The main differences for this work are that the \citet{ploeckinger_schaye_2020} tables were made with a newer {\sc CLOUDY} version \citep[v17.01,][]{cloudy_2017}, including updated atomic data, and that these tables were calculated assuming a \citet{FG20} UV/X-ray background\footnote{\citet{ploeckinger_schaye_2020} modified this background somewhat for redshifts $z > 3$, but we do not consider such high redshifts in this work.}. 
 
For the \citet{ploeckinger_schaye_2020} table lines, we look up the transitions for H-like and He-like species in the {\sc Cloudy} documentation like we did for the other tables; the wavelengths of these lines very closely match those of the earlier version. For the Fe L-shell lines, the transitions selected for the two table sets are different. The atomic data for these lines are more uncertain \citep[e.g.,][]{gu_chen_etal_2007, de-plaa_zhuravleva_etal_2012, bernitt_brown_etal_2012, wu_gao_2019, gu_raassen_etal_2019}, so slightly different wavelengths and other atomic data in different transition databases and {\sc Cloudy} versions are to be expected. We list the Fe L-shell lines we study in this work in Table~\ref{tab:linesPS20}.

For these L-shell lines, {\sc Cloudy}~17.01 uses the {\sc Chianti} database\footnote{\url{https://www.chiantidatabase.org/}} data \citep[version~7.1.4;][]{dere_landi_etal_1997, landi_young_etal_2013} by default. These default settings were used in \citet{ploeckinger_schaye_2020}. We looked up the transitions using the version~7.0 line list provided on the {\sc Chianti} website. We also checked the H- and He-like transition identifications against this list, and found they matched.

However, there was an issue with the Fe L-shell line identifications. The \citet{mewe_gronenschild_1981} wavelength and transition identification combinations do match quite well for the \ion{Fe}{xvii} 17.10~$\Ang$ and 15.26~$\Ang$ lines, and the \ion{Fe}{xviii} 16.07~$\Ang$ line. The transitions causing the \ion{Fe}{xvii} 16.78~$\Ang$ and 17.05~$\Ang$ lines are, however, reversed between the \citet{mewe_gronenschild_1981} and {\sc Chianti}~v7.0 line lists. The NIST database\footnote{accessed on 2021-03-25} \citep{gordon_hobby_peacock_1980, hutcheon_pye_evans_1976} seems to agree with the \citet{mewe_gronenschild_1981} classifications, while the lastest {\sc Chianti} database \citep[version~10.0.1][]{dere_landi_etal_1997, del-zanna_dere_etal_2021} agrees with the earlier Chianti version. There may therefore be errors in the classification of the Fe L-shell lines, and it is not entirely clear whether the {\sc Cloudy}~v7.02 \ion{Fe}{xvii} 17.05~$\Ang$ line is `the same' as the {\sc Cloudy}~v17.01 \ion{Fe}{xvii} 17.0510~$\Ang$ line.  

The transition probabilities from the NIST and {\sc Chianti} (version 7.1.4 or 10.0.1) databases do not match exactly, but they match the same-wavelength lines better than the ones identified with the same transitions. Therefore, when comparing the results from the two sets of tables we use, we will assume that the two 17.05~$\Ang$ lines are `the same'. 

We compared the surface brightness profiles we obtained for the K-$\alpha$ and He-$\alpha$-like lines using the \citet{bertone_schaye_etal_2010} tables (Fig.~\ref{fig:medmeanprof}) to those obtained for the same lines using the \citet{ploeckinger_schaye_2020} tables. The emissivity of these lines differs little in CIE, but there are larger differences in PIE, likely resulting from the different UV/X-ray backgrounds assumed in the two sets of tables. 
The resulting surface brightness profiles for the K-$\alpha$ and He-$\alpha$-like lines differ by $\lesssim 0.1$~dex where the emission is potentially observable (surface brightness $\gtrsim 10^{-2}$~photons~$\txn{cm}^{-2}\txn{s}^{-1}\txn{sr}^{-1}$, Table~\ref{tab:minsb}). These differences are often larger at lower surface brightnesses, where emission from photo-ionized gas dominates. Differences are slightly larger ($\approx 0.2$~dex) for the median surface brightnesses of the \ion{Mg}{xii} and \ion{Si}{xiii} lines in the centres of $\Mvir = 10^{12.5}$--$10^{13.0} \Msun$ haloes and of the \ion{Ne}{X} line in the centres of $\Mvir = 10^{12.0}$--$10^{12.5} \Msun$ haloes.

\subsubsection{Line luminosities}

To calculate the line luminosity for each gas (SPH) particle, we use tables which tabulate gas luminosity. The \citet{bertone_schaye_etal_2010} tables list $\log_{10} \, \mathrm{L} \, \mathrm{V}^{-1} \mathrm{n}_{\mathrm{H}}^{-2}$, the luminosity per unit volume and squared hydrogen number density, as a function of $\log_{10} \, \mathrm{T}$, $\log_{10} \, \mathrm{n}_{\mathrm{H}}$, and $z$, where T is the temperature and $z$ the redshift. The $\mathrm{n}_{\mathrm{H}}^2$ factor accounts for the first-order dependence of emission on the collision (and therefore, excitation) rate, which scales as $\mathrm{L} \, \mathrm{V}^{-1} \propto \mathrm{n}_{\mathrm{ion}} \, \mathrm{n}_{\mathrm{e}}$ in collisional ionization equilibrium (CIE), where $\mathrm{n}_{\mathrm{ion}}$ and $\mathrm{n}_{\mathrm{e}}$ are the number densities of the ions producing the line and electrons, respectively. This emission is interpolated linearly, in log space where applicable. For each particle, we multiply the table values by $\mathrm{n}_{\mathrm{H}}^{2}$ and volume (mass divided by density) to get the line luminosity.

The \citet{ploeckinger_schaye_2020} tables list $\log_{10} \, \mathrm{L} \, \mathrm{V}^{-1}$, without the first-order dependence on hydrogen number density scaled out, and these values are a function of the total metallicity $\log_{10} Z / Z_{\odot}$. Again, we interpolate these tables linearly. We multiply by the particle volume (mass over SPH density) to obtain the SPH particle luminosity. 

The line emission also depends on the abundance of the species producing the line. This dependence is linear to high accuracy: twice as many atoms of a given element mean double the number of the ions responsible for the emission, and each ion will experience the same number of excitations. (Metal ions and the electrons from these metals only make small contributions to the excitation rates.) Therefore, we scale the emission of each SPH particle by the ratio of the particle's element abundance to the solar abundance that the tables were made for \citep[tables from][]{bertone_schaye_etal_2010} or by the ratio of the SPH particle's element abundance to the assumed element abundance at the particle's total metallicity \citep[tables from][]{ploeckinger_schaye_2020}. 

Note that the solar and element abundances assumed in the two sets of tables are different. The element abundances in the {\eagle} simulations do not depend on this choice, and we calculate the emission by scaling the emission values from each table by the appropriate table value. 
We list the solar abundances used in the \citet{bertone_schaye_etal_2010} table generation in Table~\ref{tab:Zsol}. These are the {\textsc CLOUDY} v7.02 \citep[last documented in][]{cloudy} default abundances. 

\begin{table}
\caption{Solar abundances for the elements we use in this work. We give metallicities for the elements we generate emission lines for, in number density relative to hydrogen (column~2) and mass fraction (column~3). The corresponding total solar metallicity value is $\mathrm{Z}_{\odot} = 0.0127$. The values were also used to generate the line emission tables \citep[{e.g.,}][table~1]{bertone_schaye_etal_2010} and the EAGLE cooling tables \citep[table~1]{wiersma_schaye_smith_2009}. 
The values are the {\textsc CLOUDY} v7.02 \citep[last documented in][]{cloudy} defaults, from \citet[][H01]{holweger_2001}, \citet[][AP01]{allendeprieto_lambert_asplund_2001}, and \citet[][AP02]{allendeprieto_lambert_asplund_2002}. }
\label{tab:Zsol}
\centering
\begin{tabular}{l r@{$\times$}l r@{$\times$}l l }
\hline
element & \multicolumn{4}{c}{metallicity} & source \\
 & \multicolumn{2}{c}{$n_{\mathrm{elt}} \,/\, n_{\mathrm{H}}$} & \multicolumn{2}{c}{$\rho_{\mathrm{elt}} \,/\, \rho_{\mathrm{tot}}$} \\
\hline
C & $2.45$&$10^{-4}$ & $2.07$&$10^{-3}$ & AP02 \\
N & $8.51$&$10^{-5}$ & $8.36$&$10^{-4}$ & H01 \\
O & $4.90$&$10^{-4}$ & $5.49$&$10^{-3}$ & AP01 \\
Ne & $1.00$&$10^{-4}$ & $1.41$&$10^{-3}$ & H01 \\
Mg & $3.47$&$10^{-5}$ & $5.91$&$10^{-4}$ & H01 \\
Si & $3.47$&$10^{-5}$ & $6.83$&$10^{-4}$ & H01 \\
Fe & $2.82$&$10^{-5}$ & $1.10$&$10^{-3}$ & H01 \\
\hline
\end{tabular}
\end{table} 

\begin{figure*}
\includegraphics[width=0.7\textwidth]{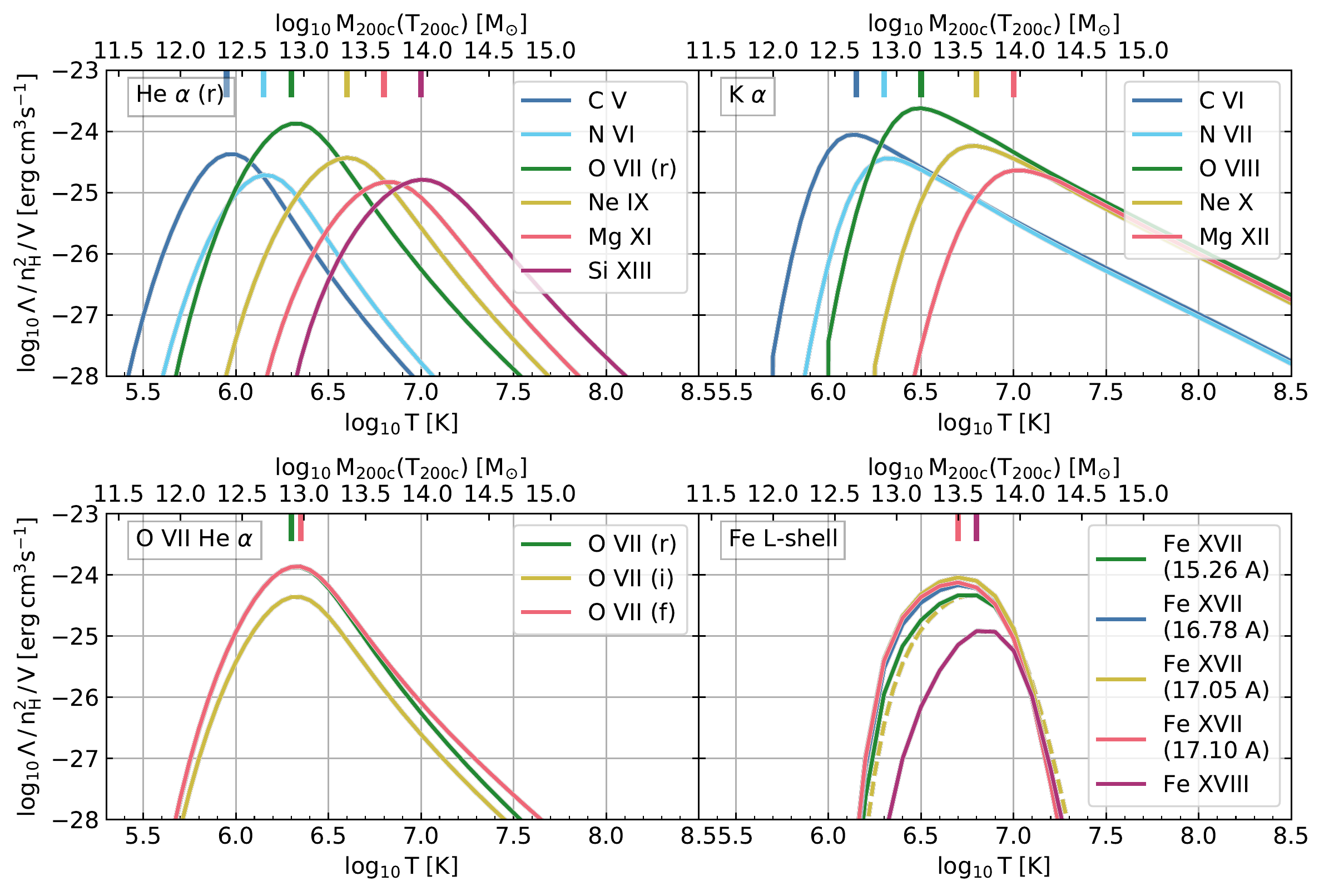}
\caption{Emissivity for CIE conditions ($\mathrm{n}_{\mathrm{H}} = 10 \pcc$, $z=0.1$) assuming solar abundances as a function of temperature for the different lines we study in this work. Vertical, coloured ticks on the top axis indicate the temperature where the emissivity of each line peaks, and the top axis shows the halo mass where the virial temperature $\Tvir$ matches the temperature on the bottom axis for a first-order prediction of which halo masses are best probed with which emission lines. We group the lines into different panels for legibility: the different He-$\alpha$-like resonance lines (top left) and K-$\alpha$ lines (top right) show curves of the same shape. Note the relatively shallow decline of the K-$\alpha$ emissivity towards high temperatures. The different \ion{O}{vii} He-$\alpha$-like lines (bottom left) have similar curve shapes, though they are not identical. The iron L-shell lines (bottom right) have more strongly peaked emissivities than the other lines. The curves for the \ion{Fe}{XVII}~16.78, 17.05, and~17.10~$\Ang$ lines largely overlap in the plot. We show the \citet{bertone_schaye_etal_2010} table \ion{Fe}{xvii}~17.05~$\Ang$ emissivity curve as a dashed, yellow line. }
\label{fig:emcurves}
\end{figure*}

We show some of the properties of these lines in  Fig.~\ref{fig:emcurves}, and in Tables~\ref{tab:lines} and~\ref{tab:linesPS20}.
Many of these lines have been explored in more detail in \citet{bertone_aguirre_schaye_2013}. Note that much of this applies to {\eagle}, even though the paper uses the OWLS simulations, because the radiative cooling model is the same. 


The line selection consists of Lyman~$\alpha$-like (K-$\alpha$) lines from H-like ions, He-$\alpha$-like (resonance) lines from He-like ions, and for iron, a number of lines from the L-shell ions. The differences between the different K-$\alpha$ and He-$\alpha$-like resonance lines in Fig.~\ref{fig:emcurves} (top right and left panels, respectively) can largely be explained by the different solar element abundances (peak heights) and element numbers (peak temperatures). 

The uncertainty in the atomic data for the Fe~L-shell lines is illustrated by the comparison between the \ion{Fe}{xvii} 17.05~$\Ang$ lines from the two table sets. Note that the curves are scaled to the same metallicities and element abundances, so differences in assumed abundances do not explain the difference. For the K~$\alpha$ en He-$\alpha$-like lines, the differences between the curves from the different table sets are $\lesssim 0.1 \dex$ around the emissivity peaks.

\subsection{Surface brightnesses}
\label{sec:sbmeth}

We calculate surface brightnesses similarly to the column densities in \citet{wijers_schaye_etal_2019, wijers_schaye_oppenheimer_2020}. First, we calculate the luminosity of each gas particle as described in \S\ref{sec:lines}, then we project the particles onto a grid, using the \citet{wendland_1995} C2 kernel as the assumed shape of each gas particle, scaled by its smoothing length. We choose a pixel size of $3.125 \ckpc$, matching that used in \citet{wijers_schaye_etal_2019, wijers_schaye_oppenheimer_2020}. 

Before projecting, we divide the particles into `slices' along the Z-axis of the simulation (an arbitrary direction relative to haloes and galaxies). Each slice is $6.25 \cMpc$ thick, again matching previous work. 
We then divide the luminosity in each pixel by its angular size and squared luminosity distance to get a surface brightness.

For surface brightness profiles, we take these maps and extract surface brightnesses at different distances to central galaxies. We average all the values in annular bins. For median profiles, we use the median, in each annular bin, of the annular averages around individual central galaxies. 
We calculated the mean profiles by similarly averaging the annular means around the different galaxies at each impact parameter.
We use impact parameter bins of $0.1 \dex$ for the medians and $0.25 \dex$ for the averages, because the average profiles are quite noisy using $0.1 \dex$ bins. The innermost bin covers the $0$--$10 \pkpc$ range. 

In our surface brightness profiles, we ignore any possible emission from the star-forming gas. Assuming the star-forming gas is at $10^{4} \K$ (warm ISM), emission from this gas is negligible.
A hot ISM phase might generate more emission, but modelling this phase would come with large uncertainties because the phases of the ISM are poorly resolved in {\eagle}. Similarly, emission from (central and satellite) galaxies, e.g., X-ray binaries or AGN, is not included in our profiles. 

Some emission from galaxies is included, however: gas that has recently been heated by supernova or AGN feedback. This gas will eventually expand and cool, but just after a feedback event, it will be hot (heated to $10^{7.5}$ or $10^{8.5}$~K, respectively) and dense, and will therefore be relatively bright in X-ray emission. However, these temperatures are high for the lines we study (Fig.~\ref{fig:emcurves}). We will later see that the line-luminosity-weighted temperature of these haloes is generally well below these high temperatures. Luminosity-weighted mean temperatures can reach $\gtrsim 10^{7}$~K in  the centres of haloes with $\Mvir \lesssim 10^{11.5}$~K, but we find that these haloes are too faint to observe even with emission from this gas included. This effect is present for the K-$\alpha$ lines and the \ion{Mg}{xi} and \ion{Si}{xiii} He-$\alpha$-like resonance lines, which have relatively high emissivities at these high temperatures. 

In Appendix~\ref{app:directfb}, we discuss the effect of this recently feedback heated gas on the surface brightness. In short, the effects are larger for the mean surface brightnesses than for the medians. Within $0.1 \Rvir$, the effect on the median profiles can be substantial if the halo virial temperature is small compared to the temperature at which the emissivity peaks. For the mean profiles, the effects can be substantial for haloes up to this virial temperature.
At these small impact parameters, we expect that, in practice, emission from the central galaxy itself would make it difficult to detect any CGM emission. At impact parameters between $0.1$ and $1 \Rvir$, the effects are typically small where the surface brightness is high enough that we would expect emission to be detectable by upcoming missions. They can be large when a halo seems to be marginally detectable, or not detectable at all.

Our surface brightness profiles therefore reflect the emission from the gas surrounding galaxies, but not from the galaxies themselves.
This is the gas we are aiming to characterize. Emission from the galaxies themselves may, in practice, make it more difficult to detect the surrounding CGM in emission.

In this work, we use SPH-smoothed element abundances ({\sc SmoothedElementAbundance} in {\eagle}) to calculate luminosities for consistency with the radiative cooling in {\eagle}. We also use these abundances to calculate metal mass fractions and hydrogen number densities.

\subsection{Galaxies and haloes}
\label{sec:mapproc}

We take galaxies and haloes from the {\eagle} public data release \citep{mcalpine_helly_etal_2016}. These are identified in {\eagle} by first finding haloes using a friends-of-friends method, 
where any particles (resolution elements) separated by less than 0.2 times the mean inter-particle distance are connected, and all connected particles define a halo. This algorithm is applied to dark matter particles, and other particles (gas, stars, and black holes) are then classified in the same way as their nearest dark matter particle. The centre of potential of the halo is the particle with the minimum gravitational potential energy. 

Galaxies were found within those haloes using {\sc subfind} \citep{springel_white_etal_2001, dolag_borgani_etal_2009}, and the central galaxy is the one containing the halo centre of potential. The {\sc subfind} code finds overdense regions within these haloes, and subhaloes are identified as the self-bound parts of these overdense regions. This binding factors in gravitational potential energy and kinetic energy, as well as thermal energy for gas.

The halo mass $\Mvir$ was determined from the centre of potential: a sphere was grown around this centre until its average internal density was $200$ times the critical density. The radius of that sphere is $\Rvir$ and the enclosed mass is $\Mvir$. 
When we calculate 2d (impact parameter) or 3d radial distances to halo centres, we use the centre of mass of the central galaxy (subhalo~0) instead of the centre of potential, to approximate the light-weighted centre of the galaxy that might be used in observations. 

The temperature $\Tvir$ of the hot CGM phase at the halo radius $\Rvir$ can be estimated from the virial theorem, assuming hydrostatic equilibrium, with pressure support coming from the thermal pressure of the hot phase: 
\begin{equation}
\label{eq:Tvir}
\Tvir = \frac{\mu \mathrm{m}_{\mathrm{H}}}{3 k}  G  \Mvir^{2/3} (200 \rho_{\mathrm{c}})^{1/3},
\end{equation}
where $\mu$ is the mean particle mass in units of the hydrogen atom mass $\mathrm{m}_{\mathrm{H}}$, $k$ is the Boltzmann constant, $G$ is Newton's constant, and $\rho_\mathrm{c}$ is the critical density. 
We assume $\mu = 0.59$, for primordial gas with fully ionized hydrogen and helium.
The assumption of hydrostatic equilibrium is not valid in especially the inner CGM of $\Lstar$ galaxies in the {\eagle} simulation
\citep{oppenheimer_2018}, but the volume-filling phase in  the $\sim \Lstar$ CGM of {\eagle} galaxies is still at X-ray producing temperatures, $\gtrsim 10^{6}$~K \citep[e.g.,][]{wijers_schaye_oppenheimer_2020}.

\subsection{CGM definitions}
\label{sec:CGMdef} 

The CGM does not have one clear definition. It is roughly the gas surrounding a central galaxy, but it does not have a clear inner or outer boundary. 
Therefore, in our 3D profiles of gas properties
we show gas over a large radial range. When calculating total luminosities in halo mass ranges (Fig.~\ref{fig:Lsplit}), we use the `FoF and gas within $\Rvir$' definition of haloes and CGM. For subhalo gas, we use the {\sc subfind} definition, where subhalo gas is gravitationally bound to a subhalo other than subhalo $0$, as indexed by {\sc subfind}. The subhalo with index 0 is the central galaxy, and all gas bound to the halo, but no specific subhalo, is attributed to subhalo 0 by {\sc subfind}. In Fig.~\ref{fig:Lw}, where we use luminosities of individual haloes, we define the CGM as all the non-star-forming gas within $\Rvir$ of (the center of mass of) the central galaxy. In Figs.~\ref{fig:Ltot} and~\ref{fig:3dprof_em} we do include star-forming gas. We note that the contribution of this star-forming gas to the halo luminosity of these lines is negligible.

\section{Detectability}
\label{sec:det}

The detectability of emission is not easy to define. For example, \citet{das_mathur_gupta_2020} use two different tests to investigate how far from the galaxy they can detect emission. Detection might be limited by backgrounds and foregrounds (both astrophysical and instrumental), and non-CGM emission correlated with the source. Such correlated emission would be, for example, X-ray emission from X-ray binaries, hot ISM, or AGN in central and satellite galaxies. Since backgrounds and foregrounds (may) vary with time and position on the sky, these need to be fit to observations along with the object of interest, meaning that systematic errors in the background (models) are also relevant.

Here, we investigate detectability with a number of instruments under somewhat optimistic assumptions: we consider only statistical errors (Poisson noise in total counts), where backgrounds contribute to the noise, but not the signal, and we ignore the effects of other X-ray sources correlated with the CGM. Modelling galaxy X-ray emission would be difficult, since {\eagle} does not resolve the multi-phase ISM or the time and spatial scales governing AGN variability, or assume binary fractions in its star formation and feedback model. However, it has been shown that e.g., aggregate AGN feedback and the resulting quenching of star formation anti-correlate with X-ray emission from the CGM at $z \approx 0$ in $\sim \Lstar$ galaxies \citep{davies_crain_etal_2019}, meaning that just `painting in' a galaxy of the right stellar mass might be insufficient. Modelling this galactic X-ray emission is beyond the scope of this paper.

We focus on the detection of individual emission lines and ignore the issue of detecting them over the continuum emission of the CGM. For high-mass systems, such as clusters, this might be an important limitation, but then possible observations of clusters in X~rays have already been studied by e.g., \citet{lotti_cea_etal_2014}, and can make use of present observations in modelling. The spectral resolution (full width at half maximum) of the X-IFU should be 2.5~eV up to 7~keV \citep{Athena_2018_07}. For rest-frame energies between 0.3 and 2~keV, this corresponds to a velocity range between $2500 \kmps$ and $375 \kmps$. These ranges are large enough that we assume the redshift of the emission line is known with sufficient precision relative to the central galaxy redshift, and that redshift trials are not an issue. (Note that this is not necessarily the case for high spectral resolution measurements with Lynx.) We ignore blending of different emission lines; in our line sample, the \ion{Fe}{xvii}~17.05 and 17.10~$\Ang$ lines would be blended. 

We assume an intrinsic line width ($b$-parameter) of $100 \kmps$ for each emission line when we calculate the signal it would produce in each instrument. This is well below the velocity resolution of XRISM Resolve, the Athena X-IFU, and the Lynx Main Array (Table~\ref{tab:ins}). These resolutions are worse than $375 \kmps$ over the line energy range we explore. Thermal line widths in the energy range producing much of the emission (Tables~\ref{tab:lines} and~\ref{tab:linesPS20}) are at most $210 \kmps$ (for \ion{Mg}{xii}). For galaxy groups, ($\Mvir > 10^{13.2} \Msun$), the halo circular velocity does exceed $375 \kmps$, so in the more massive systems we investigate, intrinsic line widths may be large enough to affect the observed line width, and hence the signal to noise ratio. This will be true across lines and halo masses for the Lynx Ultra-High Resolution array given its very high spectral resolution (Table~\ref{tab:ins}).

This means that for lower-mass haloes, we expect the line width to be determined primarily by the spectral resolution of XRISM Resolve, the Athena X-IFU, and the Lynx Main Array. This means that the precise assumptions we make about the intrinsic line width will not be very important in the determination of the detection limits at these halo masses.
At halo masses $\gtrsim 10^{13} \Msun$, this will no longer be true. 
However, we find that emission from these high-mass haloes is typically quite comfortably detectable with Athena and both Lynx instruments, and that the impact parameter where the emission crosses the detectability threshold is in a range where the surface brightness declines rapidly (Fig.~\ref{fig:medmeanprof}). This means that the extent of detectable emission for these haloes will be relatively insensitive to the precise detectability limit. Therefore, we expect that our detectability estimates for XRISM Resolve, the Athena X-IFU, and the Lynx Main Array are not very sensitive to the assumptions we make about the line width. The effects might be larger for the Lynx Ultra-High Resolution Array, which will likely spectrally resolve emission lines across halo masses.

Finally, to get a single surface brightness limit, we assume uniform emission, at least within each region the surface brightness is measured in.
This brings us to the following equations, adapted from \citet{takei_ursino_etal_2011} to include instrumental backgrounds:
\begin{equation}
\txn{r} = \int_{\txn{line}} \txn{d} \txn{E} \sum_{j} \txn{SB}(\txn{E}) \, \txn{A}_{\txn{eff}}(\txn{E}) \, \txn{LSF}(\txn{E}, j)  
\label{eq:countrate}
\end{equation}
where $\txn{r}$ is the count rate per unit angular size, E is the energy, $j$ is the spectral channel, SB is the surface brightness, $\txn{A}_{\txn{eff}}$ is the effective area, and LSF is the line spread function. The sum over the channels would be centred on the channel corresponding to the emission line energy; the range of extraction can be varied. This describes the conversion from photons to counts as encoded in the instrument response files. 
Then
\begin{equation}
N_{\sigma} = \frac{ \txn{r}_{\txn{line}}}{\sqrt{ \txn{r}_{\txn{line}}+ \txn{r}_{\txn{bkg}} }} \sqrt{\txn{t}_{\txn{exp}} \, \Delta \Omega},
\label{eq:nsigma}
\end{equation}
where $N_{\sigma}$ is the detection significance in units of $\sigma$, 
$ \txn{r}_{\txn{line}}$ and $\txn{r}_{\txn{bkg}}$ are the count rates per unit observed solid angle for the line and total background, respectively,
$\Delta \Omega$ is the angular size of the observed region, 
and $\txn{t}_{\txn{exp}}$ is the exposure time. This assumes that errors in background modelling are negligible. Finally, we assume $N_{\sigma} \geq 5$ constitutes a detection.

Given the response functions and backgrounds, we can therefore estimate what line surface brightness would be detectable. Note that this surface brightness is not the intrinsic surface brightness of the source. Especially for the lower-energy lines we consider, Galactic absorption will reduce the amount of light that makes it to the instrument. Since this is a very simple correction, we give absorbed minimum detectable surface brightnesses and unabsorbed minima assuming the same Galactic absorption as in the X-IFU background model  \citep{maccammon_almy_etal_2002}: an \code{xspec} \code{wabs} model \citep{morrison_mccammon_1983} with a hydrogen column density $\txn{N}_{\txn{H}} = 1.8 \times 10^{20} \pcmsq$ (model parameter value 0.018). 

Nominally, we assume an extraction area for surface brightnesses of $\Delta \Omega = 1 \txn{arcmin}^2$. This corresponds to a circle of radius $64 \pkpc$ at redshift~0.1. We try exposure times of 0.1, 1, and 10~Ms. Because the extraction area and exposure time are degenerate for the purposes of minimum surface brightness estimates, we report their product instead of individual values. 
We generally find that 100~ks and $1 \us \txn{arcmin}^2$ would not be sufficient to detect line emission, so we focus on the larger values. 

\subsection{Overview of instruments}

For the Athena X-IFU, we used the response matrices and backgrounds provided on the X-IFU website \footnote{\url{http://x-ifu.irap.omp.eu/resources-for-users-and-x-ifu-consortium-members/\#accordion-item-latest-x-ifu-response-matrices}, downloaded on September 28, 2020.}. \citet{athena_ifu_2016, Athena_2018_07}, and \citet{athena_xifu_2014} describe the production of the response matrices. The responses assume the cost-constrained configuration, in the baseline filter configuration (open filter wheel position). \citet{lotti_pernati_etal_2012, lotti_cea_etal_2014} document the instrumental background estimates, and the \citet{maccammon_almy_etal_2002} astrophysical background (applicable to sources at high galactic latitudes) is used. For the AGN contribution to the background, it is assumed that 80~per~cent of the AGN can be (spatially) resolved and removed.

For the Science with the X-ray Imaging and Spectroscopy Mission \citep[XRISM;][]{xrism_whitepaper_2020}, we consider the Resolve instrument. We use the response files and instrumental backgrounds from the XRISM online database\footnote{\url{https://heasarc.gsfc.nasa.gov/FTP/xrism/prelaunch/simulation/sim2/}, downloaded on October 20, 2020.}. We use the models for a 5~eV spectral resolution (FWHM), with the gate-valve open. We use the \code{.arf} file for a constant surface brightness disk with a radius of 5~arcmin. 

For the astrophysical background, we use a model from \citet{simioniescu_werner_etal_2013}, fit to Suzaku and ROSAT data taken around the Coma cluster (but in regions free from cluster emission), which was the target of the study. This includes the AGN background. Given the size of the PSF (1.2$\arcmin$) compared to the field of view (2.9$\arcmin$) of the Resolve instrument, we do not expect excising this point source background will be feasible. There might be similar issues separating galactic emission from CGM emission here, especially a possible hot ISM contribution, since this could have a very similar spectrum (collisionally ionized plasma) as the warm/hot CGM. Given a sufficiently deep galaxy survey, it might be possible to avoid this confusion by targeting a galaxy-free region of the CGM. At our nominal redshift ($0.1$), it will likely not be possible to avoid satellite galaxies given the extent of the point spread function (Table~\ref{tab:ins}).

For Lynx, we use data provided by Alexey Vikhlinin (private communication, 2020). We investigate detectability with the  \code{.rmf} response files provided by Alexey Vikhlinin and \code{.arf} files downloaded from the SOXS instrument simulator website\footnote{\url{https://hea-www.cfa.harvard.edu/soxs/responses.html}, downloaded November 21, 2018.}. We used astrophysical and instrumental background data provided by Alexey Vikhlinin, matching the specifications in \citet{lynx_2018_08}. This means the \citet{hickox_markevitch_2007} astrophysical background model (derived from Chandra measurements) is used, and instrumental backgrounds for the Lynx X-ray Microcalorimeter (LXM) are based on Athena X-IFU predictions. Point sources are assumed to be fully resolved, and therefore subtractable from the data, in deep exposures.  

For the LXM we consider two modes \citep{lynx_2018_08}: the Ultra-High Resolution Array (0.3~eV resolution, 1~arcmin FOV) and the Main Array (3~eV resolution over a 5~arcmin FOV). The UHRA is not sensitive at higher energies ($> 0.95$~keV), so it can only be used for the lower-energy lines we investigate. 

\begin{figure*}
\centering
\includegraphics[width=0.49\textwidth]{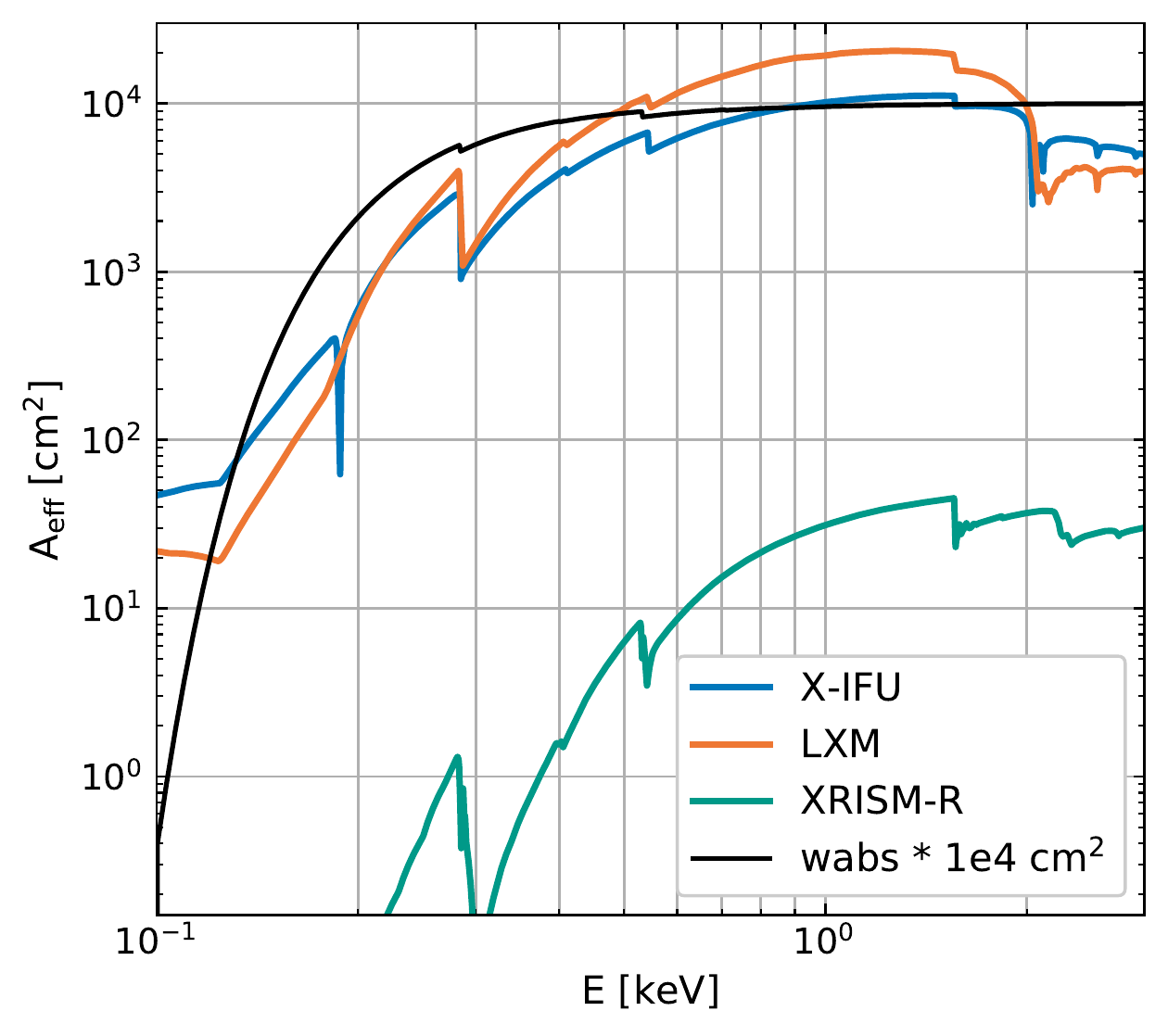}
\hfill
\includegraphics[width=0.49\textwidth]{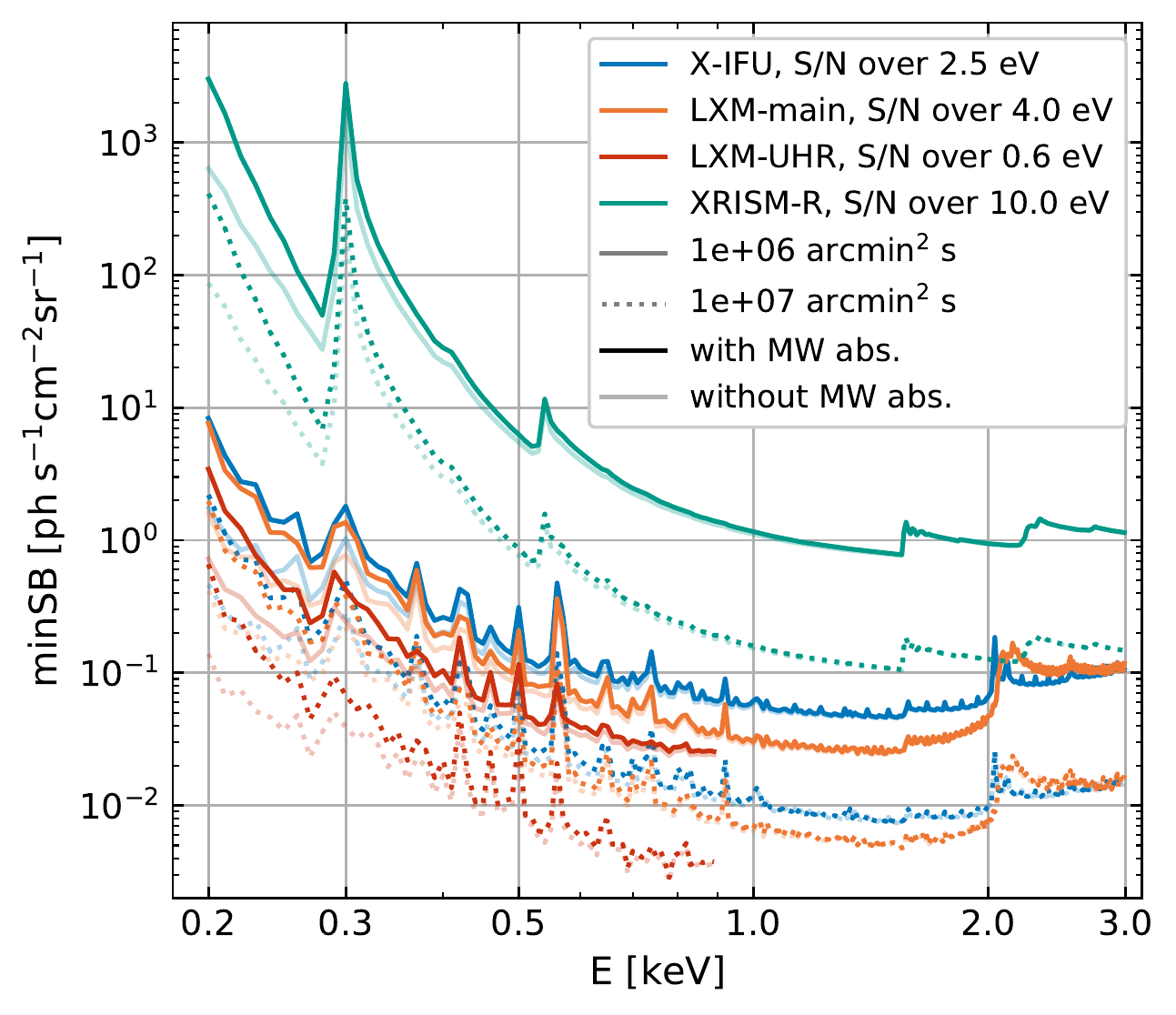}
\caption{Effective area (left) and minimum detectable surface brightness (with 5$\sigma$ confidence; right) as a function of observed line energy for the different instruments (different colours). We show the Athena X-IFU (X-IFU), the Main Array (main) and Ultra-High Resolution Array (UHR) of the Lynx X-ray Microcalorimeter (LXM), and the XRISM Resolve instrument (XRISM-R). Along with the effective area curves, we also show the fraction of transmitted radiation through the Milky Way according to the \code{wabs} model with a hydrogen column density $\mathrm{N}_{\mathrm{H}} = 1.8 \times 10^{20} \pcmsq$. We show minimum surface brightnesses for two values of exposure time and angular extent of the source: $\mathrm{t}_{\mathrm{exp}} \, \Delta \Omega = 10^{6}$ and $10^7 \us \mathrm{arcmin}^{2} \, \mathrm{s}$, with solid and dotted lines, respectively. Saturated and faded colours show minimum surface brightnesses before and after absorption by the Milky Way halo, respectively, using the \code{wabs} model shown in the left-hand panel. For each instrument, we indicate the (full) energy range, centred on the line energy, over which the signal and noise were measured. We assume an intrinsic line width of $100 \kmps$.}
\label{fig:ins}
\end{figure*}

We provide an overview of these instruments, which are potentially interesting for soft X-ray emission line detections, in Table~\ref{tab:ins}. We list the point spread function (PSF) and field of view (FOV) sizes, and the spectral resolution. We show the effective area as a function of energy in the left-hand panel of Fig.~\ref{fig:ins}. From eq.~\ref{eq:countrate}, the effective area is the main factor determining the emission line counts, while the spectral resolution determines how much of the background comes with it. The point spread function helps determine background levels through the ability to resolve and remove individual background AGN, and the FOV determines how many pointings it would take to image a source. (We do not account for this in our exposure times; these are always for single pointings.) 

The PSF also determines the level of galaxy `contamination' in the CGM images. Because we do not model this galactic emission, we cannot make precise estimates, but the relative sizes of the PSFs of the instruments should at least give an idea of the relative effects.  

\begin{table*}
\caption{Basic specifications for a number of instruments, and some relevant values for $z=0.1$. We report the point spread function (PSF) and field of view (FOV) for the instruments, and translate those values to physical sizes at $z=0.1$. The PSF types are the half-power diameter (HPD) and half-energy width (HEW). For the spectral resolution, we report the full width at half maximum (FWHM). The Lynx instruments are the Main Array (MA) and Ultra-High Resolution Array (UHRA) of the Lynx X-ray Microcalorimeter (LXM). For XRISM, we report the 1.2' angular resolution achieved by Hitomi, rather than the 1.7' requirement, and similarly, Hitomi's 5.0~eV spectral resolution rather than the 7~eV requirement. The Athena X-IFU detector is hexagonal, not square; the given field of view is an equivalent diameter.}
\label{tab:ins}
\begin{tabular}{l  l l l l l l l l l}
\hline
instrument &	\multicolumn{3}{c}{PSF} 	& \multicolumn{2}{c}{FOV}		& $\Delta \mathrm{E}$ (FWHM)	& source\\
&				arcsec	& pkpc &	type 	& arcmin	& pkpc  &	eV & \\
\hline
Athena X-IFU &		5		& 10			& HEW	& 5		& 570		   & 2.5 & \citet{Athena_2017_11} \\
XRISM Resolve &	72		& 140		& HPD	& 2.9		& 330		   & 5.0 & \citet{xrism_whitepaper_2020} \\
LXM (MA)&		0.5		& 1.0			& HPD	& 5		& 570		   & 3 &  \citet{lynx_2018_08} \\
LXM (UHRA)&		0.5		& 1.0			& HPD	& 1		& 110		   & 0.3 &  \citet{lynx_2018_08} \\
\hline
\end{tabular}
\end{table*}

\begin{table*}
\caption{Minimum source surface brightness ($\log_{10}$ photons~$\mathrm{cm}^{-2} \us  \mathrm{s}^{-1} \us \mathrm{sr}^{-1}$) detectable at a $5 \sigma$ significance for the different lines and instruments at $z=0.1$, given different exposure times and angular sizes. These include the effect of Galactic absorption according to a \code{wabs} model. The limits are averaged over 11 redshifts from $z=0.095$ to $0.105$. Dashes indicate lines outside the sensitivity range of an instrument. The final column indicates the (log) difference between the minimum detectable source surface brightnesses including and excluding the effect of Milky Way absorption (\code{wabs} model with $\mathrm{N}_{\mathrm{H}} = 1.8 \times 10^{20} \pcmsq$).}
\label{tab:minsb}
\begin{tabular}{lrrrrrrrrrrrrr}
\hline
instrument & \multicolumn{3}{c}{XRISM Resolve} & \multicolumn{3}{c}{Athena X-IFU} & \multicolumn{3}{c}{LXM UHR} & \multicolumn{3}{c}{LXM main} & \multicolumn{1}{c}{$\Delta_{\mathrm{wabs}}$} \\
$\mathrm{t}_{\mathrm{exp}} \, \Delta \Omega \; [\mathrm{arcmin}^2 \mathrm{s}]$ & 1e7 & 1e6 & 1e5 & 1e7 & 1e6 & 1e5 & 1e7 & 1e6 & 1e5 & 1e7 & 1e6 & 1e5 &  \multicolumn{1}{c}{$[\log_{10} \mathrm{SB}]$} \\
\hline
C V & 0.8 & 1.7 & 2.7 & -0.7 & -0.1 & 0.6 & -1.2 & -0.6 & 0.2 & -0.8 & -0.2 & 0.4 & 0.26 \\
C VI & 1.3 & 2.2 & 3.1 & -1.0 & -0.3 & 0.5 & -1.5 & -0.7 & 0.3 & -1.0 & -0.4 & 0.4 & 0.18 \\
N VI & 0.6 & 1.5 & 2.5 & -1.2 & -0.6 & 0.2 & -1.8 & -1.0 & -0.1 & -1.3 & -0.7 & 0.0 & 0.11 \\
N VII & 0.2 & 1.0 & 2.0 & -1.4 & -0.8 & 0.1 & -2.0 & -1.2 & -0.3 & -1.6 & -0.9 & -0.2 & 0.08 \\
O VII (f) & -0.1 & 0.7 & 1.7 & -1.6 & -0.9 & -0.1 & -2.1 & -1.3 & -0.4 & -1.7 & -1.1 & -0.3 & 0.05 \\
O VII (i) & -0.1 & 0.7 & 1.7 & -1.6 & -0.9 & -0.1 & -2.1 & -1.4 & -0.4 & -1.7 & -1.1 & -0.3 & 0.05 \\
O VII (r) & -0.2 & 0.7 & 1.7 & -1.6 & -0.9 & -0.1 & -2.1 & -1.3 & -0.4 & -1.7 & -1.1 & -0.3 & 0.05 \\
O VIII & -0.2 & 0.7 & 1.7 & -1.5 & -0.9 & 0.0 & -2.1 & -1.4 & -0.4 & -1.7 & -1.1 & -0.3 & 0.06 \\
Fe XVII (17.05 A) & -0.4 & 0.5 & 1.5 & -1.7 & -1.0 & -0.1 & -2.3 & -1.5 & -0.5 & -1.7 & -1.2 & -0.4 & 0.04 \\
Fe XVII (15.26 A) & -0.5 & 0.3 & 1.3 & -1.6 & -0.9 & -0.2 & -2.2 & -1.5 & -0.6 & -1.8 & -1.2 & -0.5 & 0.04 \\
Fe XVII (16.78 A) & -0.4 & 0.5 & 1.4 & -1.8 & -1.1 & -0.2 & -2.3 & -1.5 & -0.5 & -1.9 & -1.3 & -0.5 & 0.04 \\
Fe XVII (17.10 A) & -0.3 & 0.5 & 1.5 & -1.6 & -1.0 & -0.1 & -2.3 & -1.5 & -0.5 & -1.6 & -1.0 & -0.4 & 0.05 \\
Fe XVIII & -0.5 & 0.4 & 1.4 & -1.7 & -1.0 & -0.2 & -2.3 & -1.5 & -0.5 & -1.8 & -1.2 & -0.5 & 0.04 \\
Ne IX & -0.7 & 0.2 & 1.2 & -1.8 & -1.1 & -0.3 & -2.4 & -1.6 & -0.6 & -2.0 & -1.4 & -0.6 & 0.03 \\
Ne X & -0.7 & 0.1 & 1.1 & -1.9 & -1.2 & -0.3 & - & - & - & -2.1 & -1.4 & -0.6 & 0.02 \\
Mg XI & -0.9 & 0.0 & 0.9 & -2.1 & -1.3 & -0.4 & - & - & - & -2.3 & -1.6 & -0.7 & 0.01 \\
Mg XII & -0.9 & -0.1 & 0.9 & -2.1 & -1.3 & -0.4 & - & - & - & -2.3 & -1.6 & -0.7 & 0.01 \\
Si XIII & -0.8 & 0.0 & 1.0 & -2.1 & -1.3 & -0.3 & - & - & - & -2.2 & -1.5 & -0.6 & 0.00 \\
\hline
\end{tabular}
\end{table*}

Fig.~\ref{fig:ins} shows the effective area of these instruments (left panel), taken from the \code{.arf} files we described above. For XRISM, this includes a correction for the assumed sizes of the source and the instrument field of view. We also show the effect of Galactic absorption (transmitted fraction, scaled to $10^{4} \us \mathrm{cm}^2$). The decreasing instrument sensitivities and strong Galactic absorption mean that lines at observed energies $\lesssim 0.3$~keV will be difficult to detect. The right panel of Fig.~\ref{fig:ins} shows the resulting minimum detectable source surface brightnesses as a function of line energy. Different colours indicate different instruments, line styles are for different exposure times, and faded lines indicate the minima after absorption by the Galaxy. 

The Lynx ultra-high resolution array is the most sensitive instrument, but has a limited energy range (up to 0.95~keV). After that, the Lynx main array is most sensitive; it has a larger energy range, fully covering the 0.3--2~keV range we are interested in, and a larger field of view (a diameter of $5\arcmin$ instead of $1\arcmin$). The Athena X-IFU will have a similar sensitivity at low energies, but the difference with the Lynx main array increases towards higher energies. The XRISM Resolve instrument has the lowest sensitivity, and the strongest sensitivity decrease towards lower energies. 

The features in the  sensitivity curves in the right panel of Fig.~\ref{fig:ins} are due to detector edges (jumps in the effective area curves; left panel) and features (emission lines) in the the astrophysical background. Around these lines, systematic errors are likely to contribute significantly to the actual error budget, so our estimated surface brightness limits are likely underestimated between $\approx 0.3$~and~1~keV. 
We have checked that the different astrophysical background models are similar, so differences between those models should not be driving the sensitivity differences between the instruments. 

The features are stronger for larger exposure times and effective areas. This is because, for small exposure times ($ \txn{t}_{\txn{exp}} \, \Delta \Omega \ll N_{\sigma}^2 \,/\, \txn{r}_{\txn{bkg}}$), the limiting factor for detecting an emission line is the number of detected source photons. When the exposure time increases, the astrophysical and instrumental backgrounds become relatively more important, as shown by solving equation~\ref{eq:nsigma} for the minimum detectable emission line count rate $\mathrm{r}_{\mathrm{line}}$:
\begin{equation}
\txn{r}_{\txn{line}} = \frac{N_{\sigma}^2}{2 \txn{t}_{\txn{exp}} \, \Delta \Omega} \left(1 + \sqrt{1 + \frac{4 \txn{t}_{\txn{exp}} \, \Delta \Omega \, \txn{r}_{\txn{bkg}}}{N_{\sigma}^2}} \right) . 
\label{eq:minsb}
\end{equation}
This means that emission lines in the astrophysical background have a stronger effect on the minimum detectable surface brightness when the exposure time is larger.

The effect of the effective area is most clearly seen for the XRISM resolve instrument. Here, the background emission line features are mostly absent because the instrument has a much smaller effective area than the others (Fig.~\ref{fig:ins}, left panel). Though the instrumental background count rate of XRISM Resolve is lower than for the other instruments, its smaller effective area means that the total background is nonetheless mostly dominated by the instrumental background rather than the astrophysical one, meaning the lines in the astrophysical background have a smaller effect. 

Finally, we see that Galactic absorption makes a considerable difference in what can be detected at the lowest energies, but the effect is small ($\lesssim 0.1 \dex$) at the higher energies ($\gtrsim 0.4$~keV) we consider.   

In Table~\ref{tab:minsb} we show the minimum detectable surface brightnesses (SB) for the different lines and instruments that we investigate. For example, according to Table~\ref{tab:minsb}, for Athena X-IFU the \ion{O}{viii} detection limit is 
$\mathrm{SB} = 10^{-0.9} \us \mathrm{photons}\, \mathrm{cm}^{-2} \us  \mathrm{s}^{-1} \us \mathrm{sr}^{-1}$ for $\mathrm{t}_{\mathrm{exp}} \, \Delta \Omega = 10^{6} \us \mathrm{s} \, \mathrm{arcmin}^{2}$. This means that for this surface brightness we require $10^{6}$~s to detect a region with angular size 1~arcmin$^2$, or $10^{5}$~s to detect a region of size 10~arcmin$^2$.
To convert these minimum detectable surface brightnesses to units of photons~$\txn{s}^{-1} \us \txn{cm}^{-1} \txn{arcmin}^{-2}$, multiply the values by $1.18 \times 10^{-7}$ (or subtract $7.07$ from the log values).

These minima include the effect of Milky Way absorption on the observed surface brightness. In the final column we show how much of a difference this absorption makes; there are small variations between the different redshifts used for one line, but these are $\lesssim 0.01 \dex$. We provide these differences to make it easier to calculate the minima with different absorption models or absorbing columns, at least to first order. 

\section{Results}
\label{sec:results}

We start by demonstrating how line emission overall relates to gas and haloes in {\eagle} (\S\ref{sec:resintro}). We then examine this emission as a function of impact parameter with surface brightness profiles (\S\ref{sec:sbprof}), and compare these surface brightnesses to rough estimates of what could be detected with various instruments. Finally, we examine emission-weighted gas temperatures, densities, and metallicities to study which gas produces the emission, and how it relates to the overall gas content of haloes (\S\ref{sec:emittinggas}).

\subsection{Line emission in relation to haloes}
\label{sec:resintro}

\begin{figure*}
\includegraphics[width=0.9\textwidth]{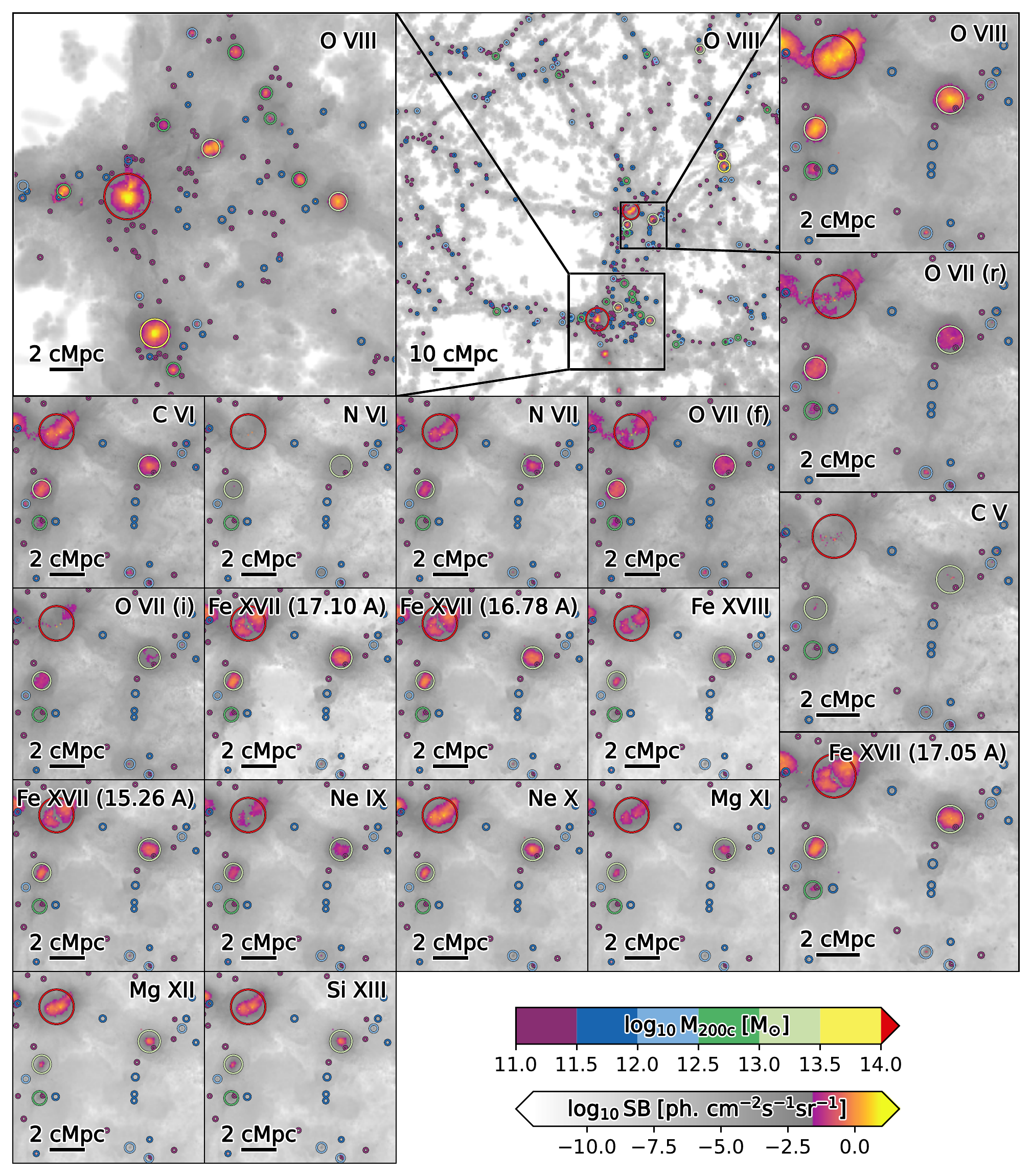} 
\caption{Surface brightness maps for a large set of X-ray emission lines. 
The \ion{O}{viii} line maps (top row) show different parts of the $100 \times 100 \times 6.25 \cMpc^{3}$ slice 
of the {\eagle} \code{RefL100N1504} volume at $z=0.1$, centred on $Z=21.875 \cMpc$. 
The projection is along the Z-axis. 
The top left panel shows a $25 \times 25 \cMpc^2$ region, centred on $\txn{X}, \txn{Y} = 57.5, 4.5 \cMpc$ at a resolution of $62.5 \ckpc$ per pixel. The middle panel shows the full slice at a resolution of $250 \ckpc$ per pixel.
Relative to the simulation coordinates, the Y coordinates were shifted by $15 \cMpc$ so the region on the left would not overlap the periodic boundary at the bottom of the image. 
The top right panel, and all the other panels, show a $12 \times 12 \cMpc^{2}$ region at $31.25 \ckpc$ per pixel, centred on $\txn{X}, \txn{Y} = 64.5, 29.5 \cMpc$.
Circles indicate the positions of haloes in the volume; these are centred on the center of mass of the central galaxy and have a radius of $\Rvir$, except for the panel showing the whole slice; there the circles indicate $2 \Rvir$. 
The color additionally shows the halo mass. The surface brightness color bar transitions to colour in the range where emission roughly becomes directly observable ($10^{-1.5} \us \mathrm{photons} \us \mathrm{s}^{-1} \mathrm{cm}^{-2} \mathrm{sr}^{-1}$). The brightest emission comes from within haloes, and the IGM emission is typically very weak. The rest of the paper uses the full {\eagle} volume.
}
\label{fig:igmstamps}
\end{figure*}

Fig.~\ref{fig:igmstamps} shows the emission from the different lines we study in a part of the $100^{3} \cMpc^{3}$ volume selected to have haloes with a range of masses. The regions we choose are among the most overdense regions in the volume. Note that this selection only applies to this figure; the conclusions and all other figures are based on the full {\eagle} volume. We indicate the positions of the haloes for comparison; the circles are at $\Rvir$ in all panels but the top, middle panel, where we indicate haloes at $2 \Rvir$. 
We can see that the line emission is brightest in haloes. 
The emission from the cosmic web is weak, and will not be directly detectable. 


In the top left of the smaller panels in  Fig.~\ref{fig:igmstamps}, we see emission from various lines that does not seem to correspond to a halo, close to the most massive halo in the panel. It also does not correspond to a halo centred just outside the region along the Z-axis. However, the emission in a number of lines reveals it is connected to that massive halo ($\Mvir = 10^{14.47} \Msun$).
In fact, the gas there 
is part of the same FoF group as the massive halo in the top left. Evidently, this is a merging system, and the top left halo is no longer separately identified. This is therefore halo gas/CGM emission, and not e.g., part of a bright filament.

\begin{figure*}
\includegraphics[width=0.7\textwidth]{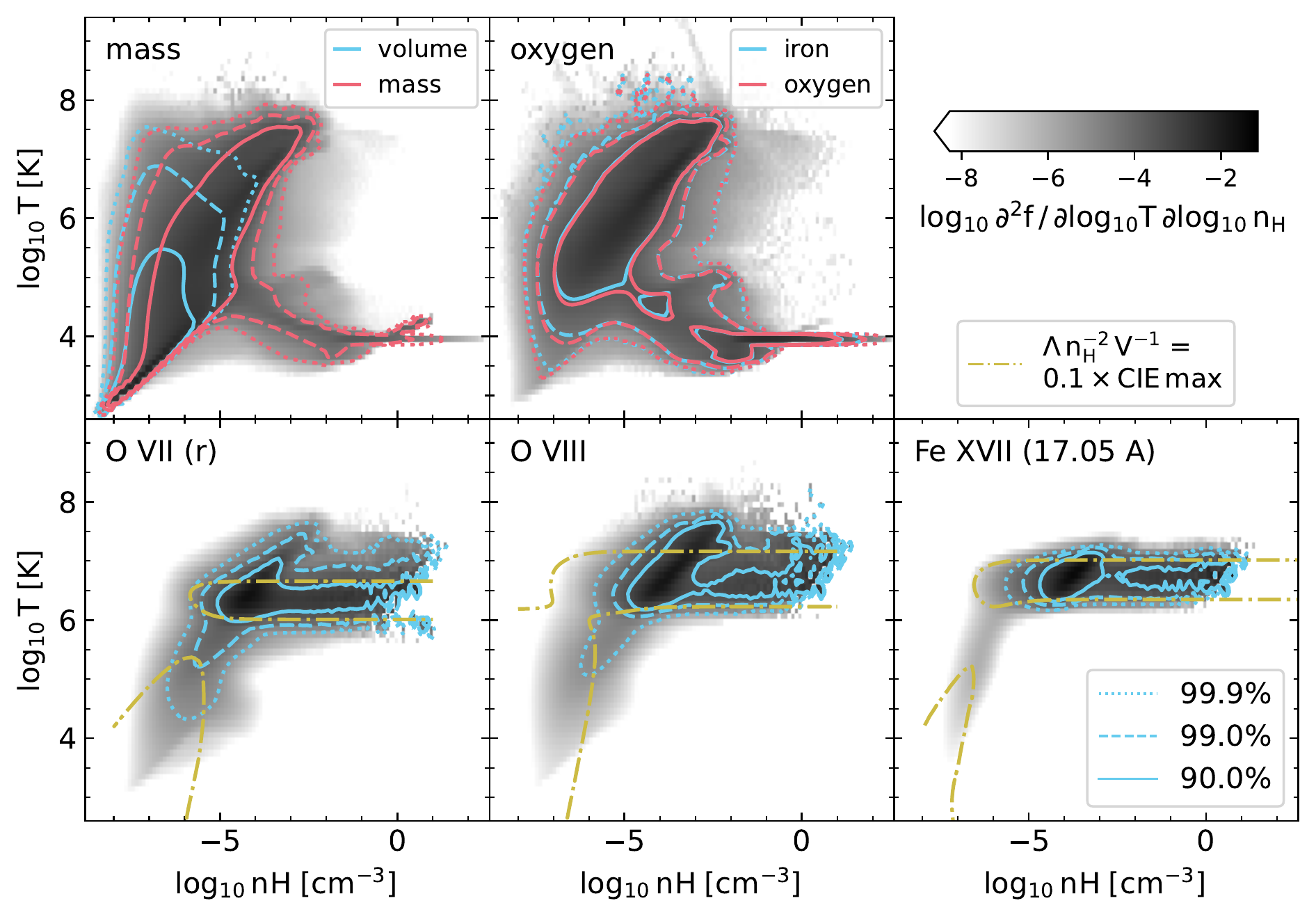} 
\caption{The distribution of mass, volume, metals, and line emission over temperature and density in the full {\eagle} $100^3$~cMpc$^{3}$ simulation volume at redshift~0.1. The top left panel shows the distribution of mass (histogram and red contours) and volume (cyan contours) in this space. The top middle panel shows the oxygen (histogram and red contours) and iron (cyan contours) distributions. The bottom panels show the distribution of (from left to right) the \ion{O}{vii}~(r), \ion{O}{viii} K-$\alpha$ and \ion{Fe}{xvii} (17.05~$\Ang$) line emission in the histograms and cyan contours. The yellow dot-dashed lines show the temperature and density range where the density-squared-normalized emissivity of the gas at fixed metallicity is at least 0.1~times as large as its maximum value in CIE. The red and cyan contours show where the enclosed fractions of the mass, volume, metals, or luminosity indicated in the legend lie in temperature and density space. The black and grey histograms show the fraction of the mass, metals, etc., per unit bin size at each temperature and density. The emission lines originate mostly from collisionally ionized gas close to the ideal temperature for each emission line, and the line emission is strongly biased towards dense gas.  
}
\label{fig:pds}
\end{figure*}

The top panels of Fig.~\ref{fig:pds} show the temperature and density of gas and metals in the $100^3$~cMpc$^{3}$ {\eagle} simulation at redshift~0.1. Gas at densities $\txn{n}_{\txn{H}} \lesssim 10^{-5} \pcc$ is mostly part of the IGM. Denser gas is mostly found in haloes, and densities $\txn{n}_{\txn{H}} \gtrsim 10^{-1} \pcc$ are typical of the ISM. Star-forming gas is included in this figure at an assumed temperature of $10^{4}$~K, and makes up the horizontal strip at high densities in the top panels. This strip intersects a population of dense gas with a temperature which increases with density. The temperature of this gas is set by the pressure floor implemented in the {\eagle} simulations. 

Most of the gas mass is in the IGM (see also Fig.~\ref{fig:Ltot}), but appreciable fractions are also in CGM and ISM. The volume is dominated by IGM.
Metals are found at many temperatures and densities. They are biased to denser gas, which is typically closer to the galaxies where the metals are formed. The different metals have very similar distributions in temperature and density, as the close overlap between the oxygen and iron contours in the top middle panel of Fig.~\ref{fig:pds} illustrates.

Line emission from these metals, on the other hand, originates almost exclusively from collisionally ionized gas, at temperatures close to the ideal temperatures for line emission in CIE. Within these constraints, emission originates from temperatures and densities where there is relatively more mass, but with a strong bias towards higher densities. That is because the luminosity of a volume of gas is proportional to its squared density. Therefore, much line emission originates in gas at densities and temperatures where gas and metals are relatively rare. 
Some emission does originate in low-density, photo-ionized gas. This fraction is generally largest for lines at lower energies and for He-$\alpha$-like lines, where photo-ionized emission can occur at higher densities than in higher-energy lines and for K-$\alpha$ and Fe~L-shell lines.

The three lines we show here are representative of a trend we see when comparing He-$\alpha$-like (\ion{O}{vii} (r)), K-$\alpha$ (\ion{O}{viii}) and Fe~L-shell (\ion{Fe}{xvii} (17.05~$\Ang$)) lines. As the width of the emissivity peak increases from Fe~L-shell to He-$\alpha$-like to K-$\alpha$ lines (Fig.~\ref{fig:emcurves}), the emission lines probe gas in a wider range of temperatures. 

\begin{figure}
\includegraphics[width=\columnwidth]{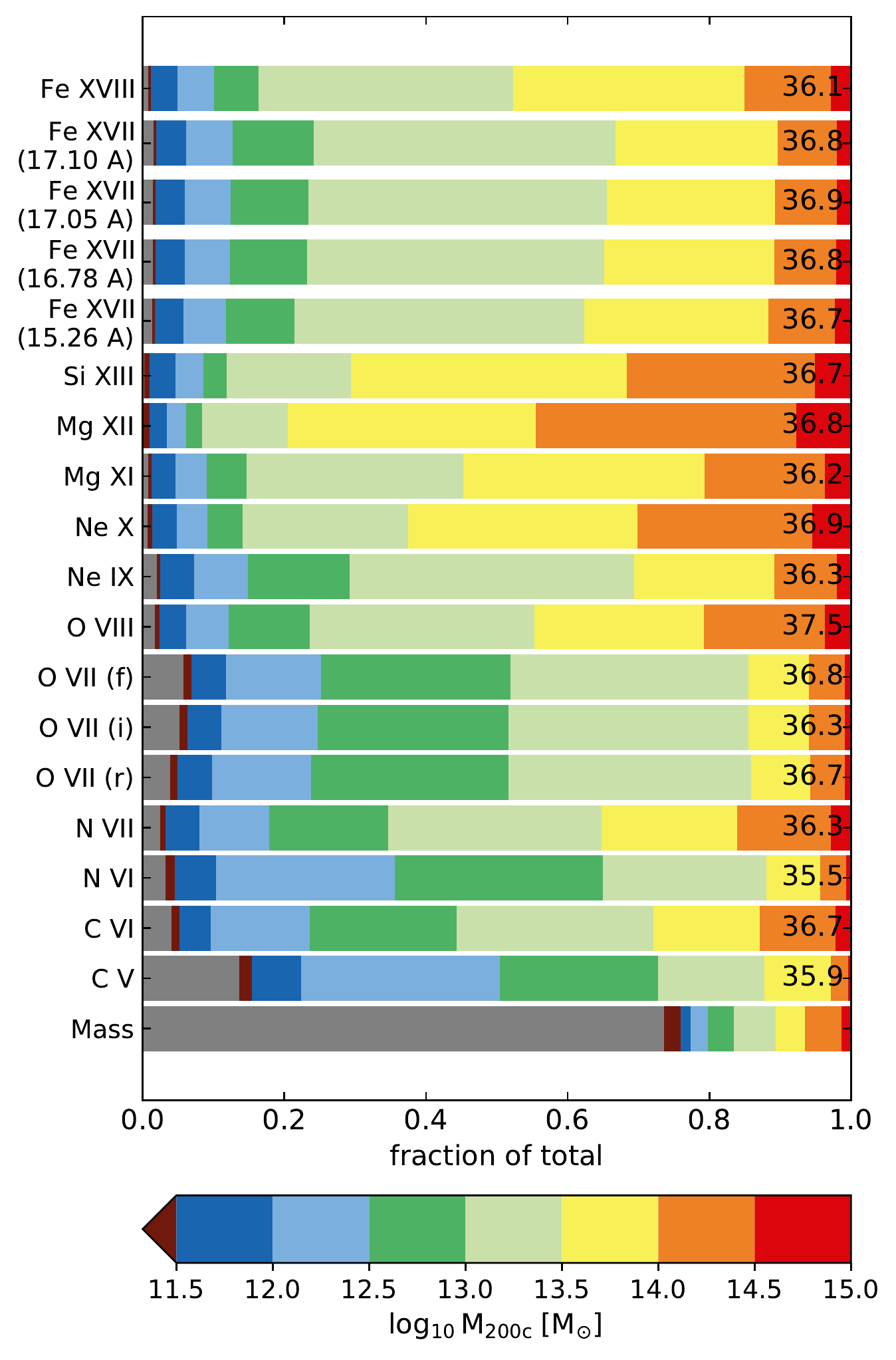}
\caption{Fraction of the total luminosity of the different lines originating in haloes of different mass (different colors) and the IGM (gray) in {\eagle} at $z=0.1$. Halo gas is anything in a FoF group or within $\Rvir$, and the IGM is anything else. The numbers at the right of each bar indicate the volume-averaged luminosity density in each line ($\log_{10} \, \mathrm{erg} \us \mathrm{s}^{-1} \cMpc^{-3}$). For comparison, we also show the fraction of the total gas mass in these components.}
\label{fig:Lsplit}
\end{figure}

Fig.~\ref{fig:Lsplit} divides the total luminosity from different lines in the {\eagle} volume into contributions from haloes of different masses and the IGM, with the mass fractions in these components shown at the bottom for comparison. Star-forming gas is included at $10^4 \K$, but its contribution to the total is negligible. Line emission is dominated by haloes, in contrast to the mass. The halo mass contributions to the emission differ between lines. The halo mass ranges contributing most to the total luminosity are mostly those for which $\Tvir$ corresponds to the CIE peak emission temperature shown in Fig.~\ref{fig:emcurves}. In \citet[fig.~2]{wijers_schaye_oppenheimer_2020}, we saw that the metals and ions producing a number of these lines are less concentrated in haloes than is the case for the emission. We expect the difference in results from the $\propto \mathrm{n}_{\txn{H}}^2$ dependence of line emission, compared to the $\propto \mathrm{n}_{\txn{H}}$ dependence of ion mass. 

Note that the IGM contributions here can differ considerably, in relative terms, between the two emissivity tables we used in the calculations, since this low-density gas is photo-ionized, and the tables assume different UV/X-ray backgrounds. The contributions to the total are small in either case.

Gas that has recently been directly heated by feedback can be responsible for a large fraction of the emission from haloes at masses where $\Tvir$ is too low for the CGM to produce detectable emission. This effect is substantial in roughly the same halo mass ranges where the effect of this gas on the surface brightness profiles of the haloes is substantial, described in Appendix~\ref{app:directfb}. This means that at low halo masses, where the contribution of a given halo mass range to the total emission of a given line is already small, it would, in general, be even smaller if the recently, directly heated gas were excluded. 

Comparing the halo emission contributions to the Fe~L-shell, He-$\alpha$-like, and K-$\alpha$ line emission, we see a secondary trend. The lines with the narrowest emissivity peaks (the iron lines, Fig.~\ref{fig:emcurves}) have most of their halo emission coming from only two halo mass bins. The He-$\alpha$-like lines, with wider emissivity peaks, have haloes over a wider mass range contributing to their total emission. The K-$\alpha$ lines tend to come from a wider range of halo masses, especially towards higher halo masses, reflecting their wider emissivity peaks, with relatively shallow slopes towards high temperatures.

We will find that many trends of line emission with halo mass are driven primarily by these two characteristics of the emissivity curves of the lines: the temperature of the emissivity peak, compared to the halo virial temperature, and the width of the peak.

\begin{figure}
\includegraphics[width=\columnwidth]{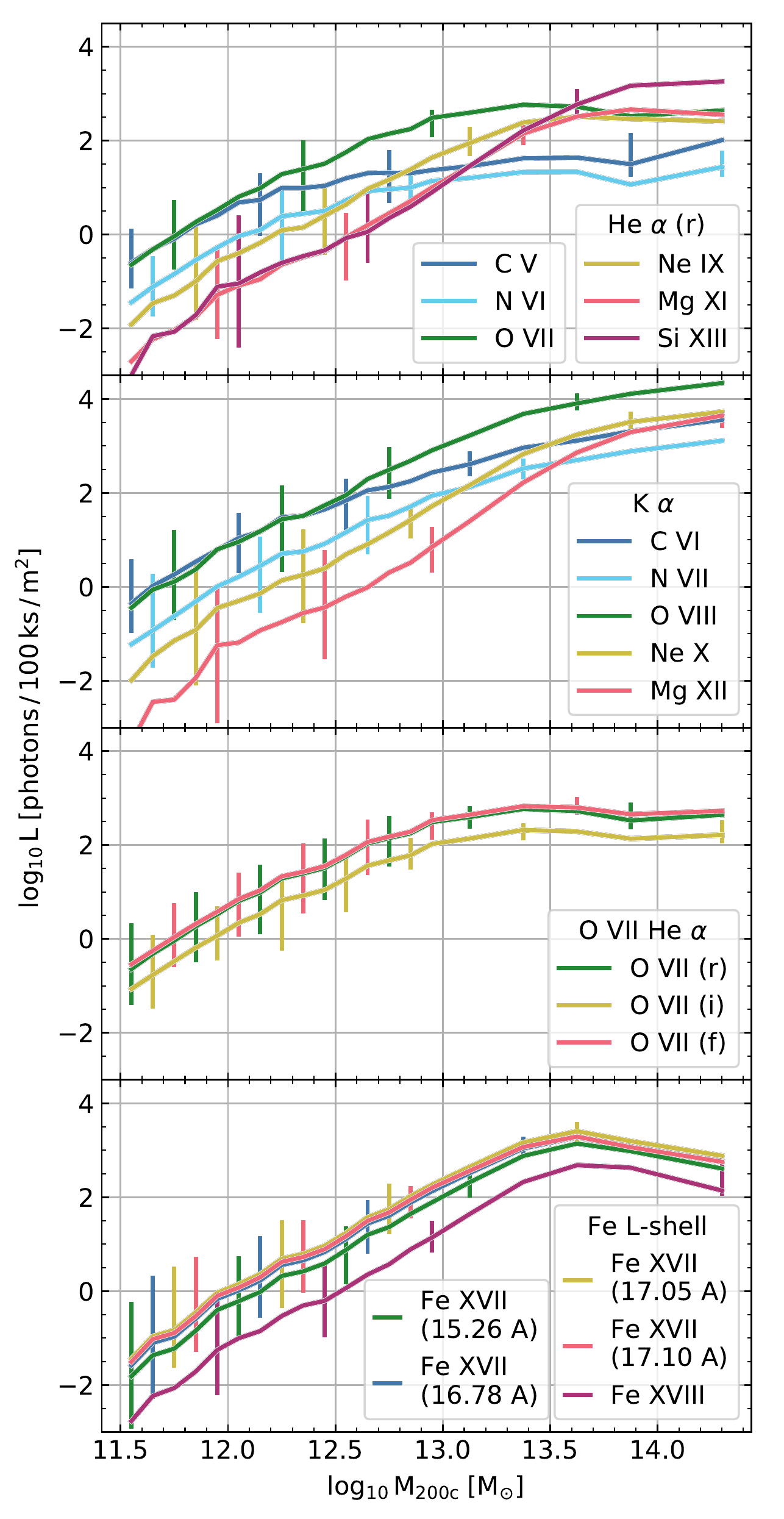}
\caption{Median flux from $z=0.1$ {\eagle} haloes (all gas within $\Rvir$) as a function of halo mass for the different lines. We show the flux in units suited for estimating photon statistics for observations: photons per 100~ks per $\txn{m}^{2}$ of effective area. We calculate the photon fluxes using a luminosity distance of $1.48 \times 10^{27} \us \mathrm{cm}$. 
Error bars, shown in alternating halo mass bins for legibility, show the central 80~per~cent of luminosities at each halo mass. (They represent scatter in the population, not noise.) Panels show, from top to bottom, the He-$\alpha$-like resonance lines, K-$\alpha$ lines, \ion{O}{vii} He-$\alpha$-like lines, and the iron L-shell lines. The \ion{O}{vii} He-$\alpha$-like resonance line is shown in both the first (top) and third panels. The curves for the \ion{O}{vii} resonance and forbidden lines largely overlap, as do the \ion{Fe}{XVII}~16.78, 17.05, and 17.10~$\Ang$ curves. Luminosities almost always increase with halo mass, and luminosity scatter is typically largest at low halo masses and low luminosities.}
\label{fig:Ltot}
\end{figure}

Fig.~\ref{fig:Ltot} shows the median line flux as a function of halo mass. Generally, we see that the oxygen lines are strongest. Though the luminosities generally increase with halo mass, we do see differences in these trends.

Fig.~\ref{fig:Ltot} also shows the scatter (central 80~per~cent) in those luminosities at fixed halo mass. This scatter is generally quite large: at least $\approx 1$~dex at $\Mvir \lesssim 10^{13} \Msun$, and $\approx 2$~dex for \ion{Ne}{x}, \ion{Mg}{xi}, \ion{Mg}{xii}, and \ion{Si}{xiii} for $\Lstar$ haloes. 
This large scatter implies that average and median surface brightnesses can differ considerably, and that the manner in which luminosity-weighted temperature, density, and metallicity distributions from different haloes are stacked can have a real impact on the resulting radial profiles. 

For the He-$\alpha$-like resonance lines in the top panel of Fig.~\ref{fig:Ltot}, we see a trend where halo luminosities tend to flatten as a function of halo mass above the emission peak temperature (Fig.~\ref{fig:emcurves}). The Fe L-shell lines even decrease in luminosity at the highest halo masses, when the haloes become hotter than their emissivity peaks. For the K-$\alpha$ lines (second panel from the top), there appears to be a weak flattening above the emissivity peak temperature halo mass.
The flattening is probably less obvious than for the He-$\alpha$-like lines because the emissivities decrease less strongly with temperature for the K-$\alpha$ lines than for the He-$\alpha$-like lines. In the third panel from the top, we see that the different \ion{O}{vii} He-$\alpha$-like lines follow very similar trends, including their scatter. 

\subsection{Surface brightness profiles}
\label{sec:sbprof}

\begin{figure*}
\includegraphics[width=0.9\textwidth]{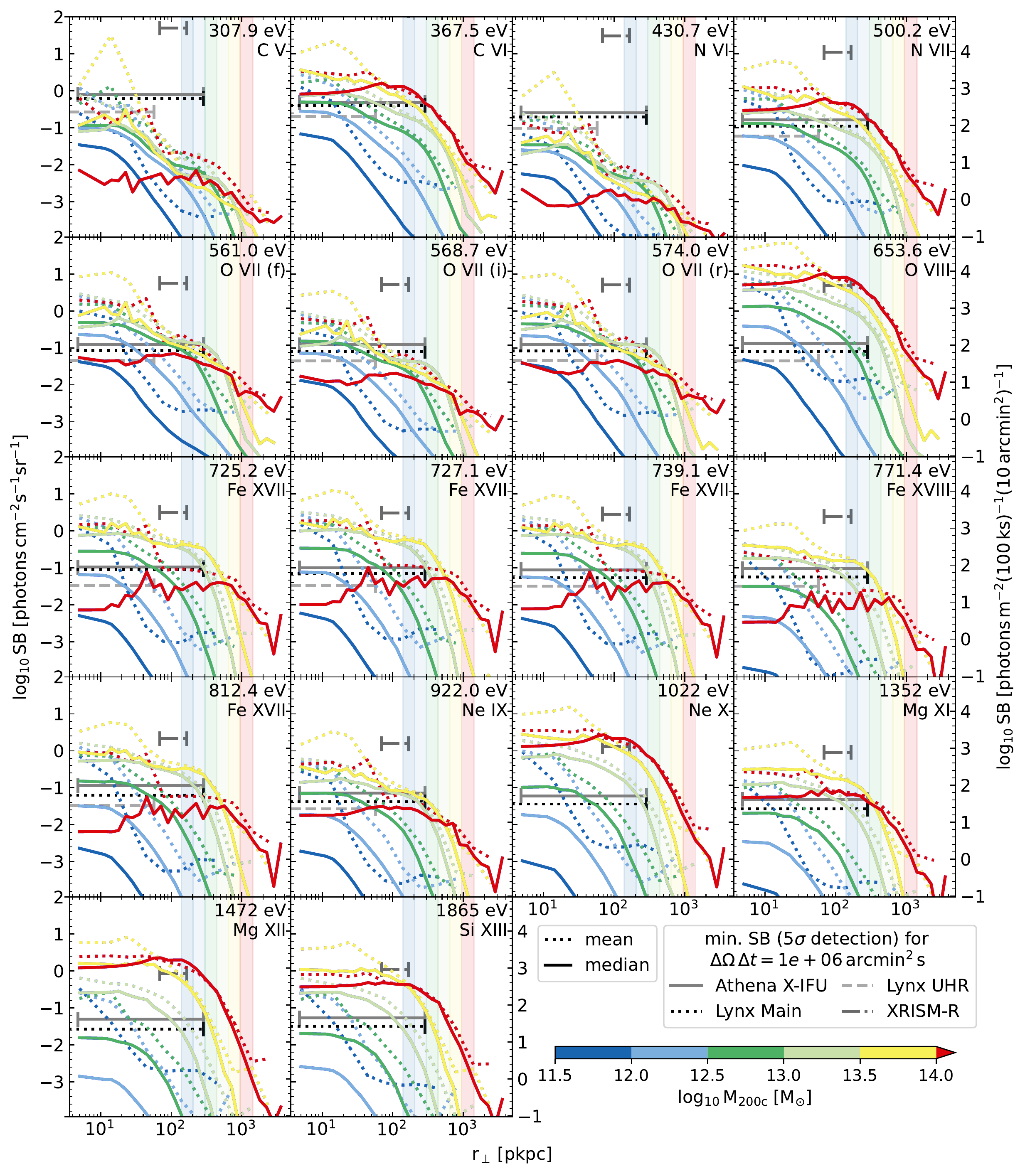}
\caption{Median and mean surface brightness as a function of impact parameter $\mathrm{r}_{\perp}$ and halo mass (colored curves). Means (dotted) are calculated using all haloes in each mass bin, or a subsample of 1000 randomly chosen haloes if the bin contains more than that. For the means, we use $0.25 \dex$ bins, starting after the first, 0--$10\pkpc$, bin. For the medians, we use $0.1 \dex$ bins. The medians are per-mass-bin medians of the annular average profiles of individual haloes. Horizontal lines in black and grey show minimum observable surface brightnesses for long integrations with different instruments. The horizontal extent of these lines indicates the point spread function (inner limit) and field of view (outer limit; half the diameters in Table~\ref{tab:ins}) of each instrument.
The emissivity curves (Fig.~\ref{fig:emcurves}) predict the halo mass for which the median SB peaks reasonably well for the He-$\alpha$-like triplets and Fe~L-shell lines, but for K-$\alpha$ lines, the emission just peaks at the highest halo mass for all lines. Comparing these curves to detection limits gives an impression of whether we could detect typical emission in these haloes. The right axis shows surface brightnesses in units useful for estimating photon numbers; for reference, $10 \us \txn{arcmin}^2$ is the area of a circle with radius $204 \pkpc$ at $z=0.1$.}
\label{fig:medmeanprof}
\end{figure*}

In Fig.~\ref{fig:medmeanprof}, we show radial surface brightness profiles. Different panels show different emission lines, and different colours correspond to different halo mass bins. The solid (dotted) lines show median (mean) profiles. For these, the starting point is a set of surface brightness maps for each emission line. We first average the surface brightnesses in annular bins around the central galaxy of each halo in a halo mass bin. Then, we take the median of these profiles in each annular bin. For the mean profiles, we average the profiles in each annular bin instead. The median profiles show the surface brightness profile we can expect for a typical halo in each mass range, while the means show what we could expect from stacking the emission in each halo mass bin. Comparing the median surface brightnesses to an instrument's detection limit indicates whether we can expect to detect about half the haloes at a given mass bin at a certain radius. Using the mean surface brightness instead yields an estimate of whether haloes in that mass bin would be detectable in a stack combining observations of many different haloes.

We use larger radial bin sizes for the means than for the medians. 
For the medians, the profiles look similar in 0.25~dex bins, but less smooth and with less sharp features. The means look noisier when using 0.1~dex bins. 
We choose these larger bins for legibility and to highlight trends with halo mass and radius. Light-coloured, vertical bands indicate the virial radii $\Rvir$ for each mass bin. Horizontal lines indicate the estimated sensitivity limits for different instruments, for an exposure times and sky area of $10^{6}$~arcmin$^2$~s. The extent of these lines indicates the point spread function and field of view of these instruments for a source at $z=0.1$.

Generally, the median surface brightnesses decline as a function of distance to the halo centre, and with the exception of the most massive haloes, typically drop off by $\gtrsim 2 \dex$ from their peak before $\Rvir$. The different behaviour for the most massive haloes ($\Mvir \geq 10^{14} \Msun$) is most likely because their virial temperatures are $\gtrsim 10^{7}$~K, above the emission peaks for all these lines, and the centres of the haloes are hotter than the outskirts (Fig.~\ref{fig:3dprof_em}). 

There is considerable scatter between haloes in the annular mean surface brightness at a given radius. We do not show this scatter to preserve legibility of the figure. The scatter is generally largest at low halo masses and in halo centres (several orders of magnitude), and smallest at large halo masses and at large impact parameters (at minimum $\approx 0.5$~dex). The 90$^{\mathrm{th}}$ percentile of the halo annular means at a given radius tends to lie close to the mean of annular means. The scatter is usually fairly symmetric about the median. 

In Appendix~\ref{app:directfb} we investigate the effect of gas recently heated directly by feedback.
In short, the effects of this gas tend to be limited to halo centres (impact parameters $\lesssim 0.1 \Rvir$) and regions where the surface brightnesses including the feedback-heated gas are too low to be detected ($\lesssim 10^{-2} \us \mathrm{photons} \, \mathrm{cm}^{-2} \mathrm{s}^{-1} \mathrm{sr}^{-1}$). The effects are larger for the mean profiles than for the medians. Where differences between mean and median profiles are relatively large, the detectability of emission in stacks (mean profiles) may hinge on the inclusion of the  gas recently heated by feedback.


Even without stacking, Athena and Lynx should be able to detect many emission lines from groups and clusters ($\Mvir \gtrsim 10^{13} \Msun$), and some lines from $\Lstar$ and local group mass haloes ($\Mvir \approx 10^{12}$--$10^{13} \Msun$) out to distances far into the CGM.
With XRISM, we should be able to detect a few emission lines from groups and clusters.
Given the declining surface brightness with impact parameter, mapping the line emission of a typical halo less massive than clusters ($\Mvir \lesssim 10^{14} \Msun$) out to $\Rvir$ will, however, be very difficult. However, it may be feasible to detect the emission statistically for a large sample of objects.

The mean profiles (dotted lines) broadly follow the trends of the medians (solid lines), with a few key differences. First, the mean surface brightnesses are larger than the medians. At halo masses close to that where the surface brightness peaks, and close to halo centres, the mean surface brightness tends to lie around the $90^{\txn{th}}$ percentile of the surface brightness distribution at each radius. However, at larger radii, differences between mean and median values become larger, and the mean is more clearly dominated by extremes. 
Generally, the differences are larger at lower median surface brightnesses, and the mean profiles often flatten out at large radii, unlike the medians. 

Additionally, there is often a stronger central peak in the mean surface brightnesses; this is typically at $\lesssim 0.1 \us \Rvir$, in a region that may still be associated with the central galaxy. Often, this bright emission is associated with recent feedback events (Appendix~\ref{app:directfb}).

At large radii, where median surface brightnesses are low, the mean profiles become noisy (especially clear when using smaller radial bins), likely dominated by bright emission in one or a few galaxies, even in mass bins with many haloes. 

The trends of the median profiles with halo mass differ between the different sets of lines (K-$\alpha$, He-$\alpha$-like, and Fe~L-shell) we investigate.
The K-$\alpha$ lines (\ion{C}{vi}, \ion{N}{vii}, \ion{O}{viii}, \ion{Ne}{x}, and \ion{Mg}{xii}) have brightnesses that consistently increase with halo mass, except in the centres of the highest mass haloes. On the other hand, the He-$\alpha$-like lines (\ion{C}{v}, \ion{N}{vi}, \ion{O}{vii} f, i, and r, \ion{Ne}{ix}, \ion{Mg}{xi}, and \ion{Si}{xiii}) have surface brightnesses that more clearly peak with halo mass. The iron L-shell lines (all the Fe lines we show) have even more extreme surface brightness peaks with halo mass. This follows the trends we saw for halo luminosities (Fig.~\ref{fig:Ltot}), except that the luminosity is spread over a larger area in more massive haloes, so roughly constant luminosities with halo mass lead to surface brightness peaks with halo mass.

In general, we find that most of these emission lines should be detectable in haloes of some mass with the Athena X-IFU. The brightest emission line is \ion{O}{viii} K-$\alpha$ (left panel in row~2 of Fig.~\ref{fig:medmeanprof}), and we expect it to be detectable in haloes of $\Mvir \gtrsim 10^{12} \Msun$ with the X-IFU. In groups and clusters ($\Mvir \gtrsim 10^{13} \Msun$), this line may even be detectable out to $\Rvir$. The other K-$\alpha$ lines are typically detectable in haloes $\gtrsim 10^{12.5}$ or $10^{13} \Msun$ with this instrument. 

Emission from the brightest He-$\alpha$-like species, \ion{O}{vii} (row 2, columns 1--3 of Fig.~\ref{fig:medmeanprof}), may also be detectable in $\Mvir \gtrsim 10^{12} \Msun$ haloes with the X-IFU.  Emission lines from \ion{C}{v} and \ion{N}{vi} (top row of Fig.~\ref{fig:medmeanprof}) will be difficult to detect at all due to the high sensitivity limits at these low energies. However, higher-energy He-$\alpha$-like lines (bottom two rows of Fig.~\ref{fig:medmeanprof}) will likely be detectable with the X-IFU in groups ($\Mvir \approx 10^{13}$--$10^{14} \Msun$), and emission from some of these lines may be detectable in $\Mvir \gtrsim 10^{14} \Msun$ or $\Mvir \approx 10^{12.5}$--$10^{13} \Msun$ haloes.

The \ion{O}{vii} forbidden line (row~2, column~1) is generally about as bright as the resonance line (row~2, column~3), and its detection prospects are similar. The intercombination line (row~2, column~2) is somewhat weaker, and would therefore be more difficult to detect. With the X-IFU and a long integration time, detecting all three lines may be possible in the inner CGM of $\Mvir \approx 10^{12.5}$--$10^{14} \Msun$ haloes, making the diagnostic information of the combination available.

The Fe~L-shell lines (rows~3 and~4) should be clearly detectable in groups ($\Mvir \approx 10^{13}$--$10^{14} \Msun$) using the X-IFU;  a few of these lines may be detectable out to almost $\Rvir$. Some Fe~L-shell emission lines from the centres of local group mass systems ($\Mvir \approx 10^{12.5}$--$10^{13} \Msun$) should also be detectable with the X-IFU. 

Overall, using the Athena X-IFU, we expect to be able to detect many different emission lines from the CGM of galaxy groups and clusters (IGrM/ICM; $\Mvir \gtrsim 10^{13} \Msun$), and a few of these lines may be detectable out to $\Rvir$. Emission from the inner CGM of $\Mvir \approx 10^{12.5}$--$10^{13} \Msun$ haloes should be clearly detectable for some of the emission lines, and marginally detectable in most of the others we study. For example, emission from a typical halo of this mass should be detectable out to $\approx 0.3 \us\Rvir$ for \ion{O}{viii} (row~2, column~4) and out to $\approx 0.1$--$0.2 \us\Rvir$ for \ion{O}{vii} (f and r, row~2) and \ion{Fe}{xvii} (725.2, 727.1, and 739.1~$\Ang$, row~3).  For $\Mvir \approx 10^{12}$--$10^{12.5} \Msun$ haloes, detections will be difficult for most emission lines we study, but detection of \ion{O}{vii} and \ion{O}{viii} emission (row~2) is likely possible with large exposure times 
($\mathrm{t}_{\mathrm{exp}} \, \Delta \Omega \approx 10^{7}$~s~arcmin$^2$).

XRISM resolve is clearly less sensitive than the X-IFU, but emission from groups and clusters ($\Mvir \gtrsim 10^{13} \Msun$) may be detectable in a few bright lines. \ion{Fe}{xvii} emission (rows~3 and~4 of Fig.~\ref{fig:medmeanprof}) may be marginally detectable with XRISM, especially if the different lines are taken together. At the spectral resolution of XRISM resolve, the 17.05 and 17.10~$\Ang$ (727.1 and 725.2~eV) Fe lines will be blended.

The Lynx MA generally has sensitivity limits similar to those of the X-IFU, but it is a bit more sensitive to line emission, especially at higher energies. The UHRA is clearly more sensitive. In addition to what is possible with the X-IFU, this instrument will enable clear detections of emission from the centres of 
 $\Mvir \approx 10^{12.5}$--$10^{13} \Msun$ haloes in more lines and out to larger impact parameters, and  it will increase the number of emission lines we can detect from $\Mvir \approx 10^{12}$--$10^{12.5} \Msun$ haloes.
 
 Note that due to the relatively small field of view of the Lynx MA, for many lines and halo masses, multiple pointings would be needed to cover the area where we expect emission to be detectable. The high spectral resolution of this instrument also likely means we underestimate some of the uncertainties involved in line detections with the UHRA. 
For example, we might have to account for at least a few different possible line centres when defining the detection significance. This can raise the $5\sigma$ surface brightness limit above the values we report here for a single redshift trial.

Despite reasonable detection prospects for a number of these haloes with the different instruments, detailed imaging will be very difficult except for the brightest lines and most massive ($\gg \Lstar$) haloes at $z=0.1$. This is because these detections would require combining large areas (at least a square arcminute), together with long exposure times (at least 1~Ms). Examining haloes at lower redshifts ($< 0.1$) might be helpful here, though too low redshifts would make the emission difficult to distinguish from local and Milky Way halo line emission.

We do not expect line emission from $\Mvir < 10^{12} \Msun$ haloes to be detectable with these instruments. Though nominally, it seems like this can be overcome by stacking in halo centres, emission here will be difficult to attribute unambiguously to the CGM (as opposed to e.g., hot ISM). Moreover, in {\eagle} this emission is largely due to gas that is at potentially unphysical temperatures and densities as a result of direct heating by the subgrid model for supernova or AGN feedback (Appendix~\ref{app:directfb}).

\subsection{Origin of the emission}
\label{sec:emittinggas}

\begin{figure*}
\includegraphics[width=0.8\textwidth]{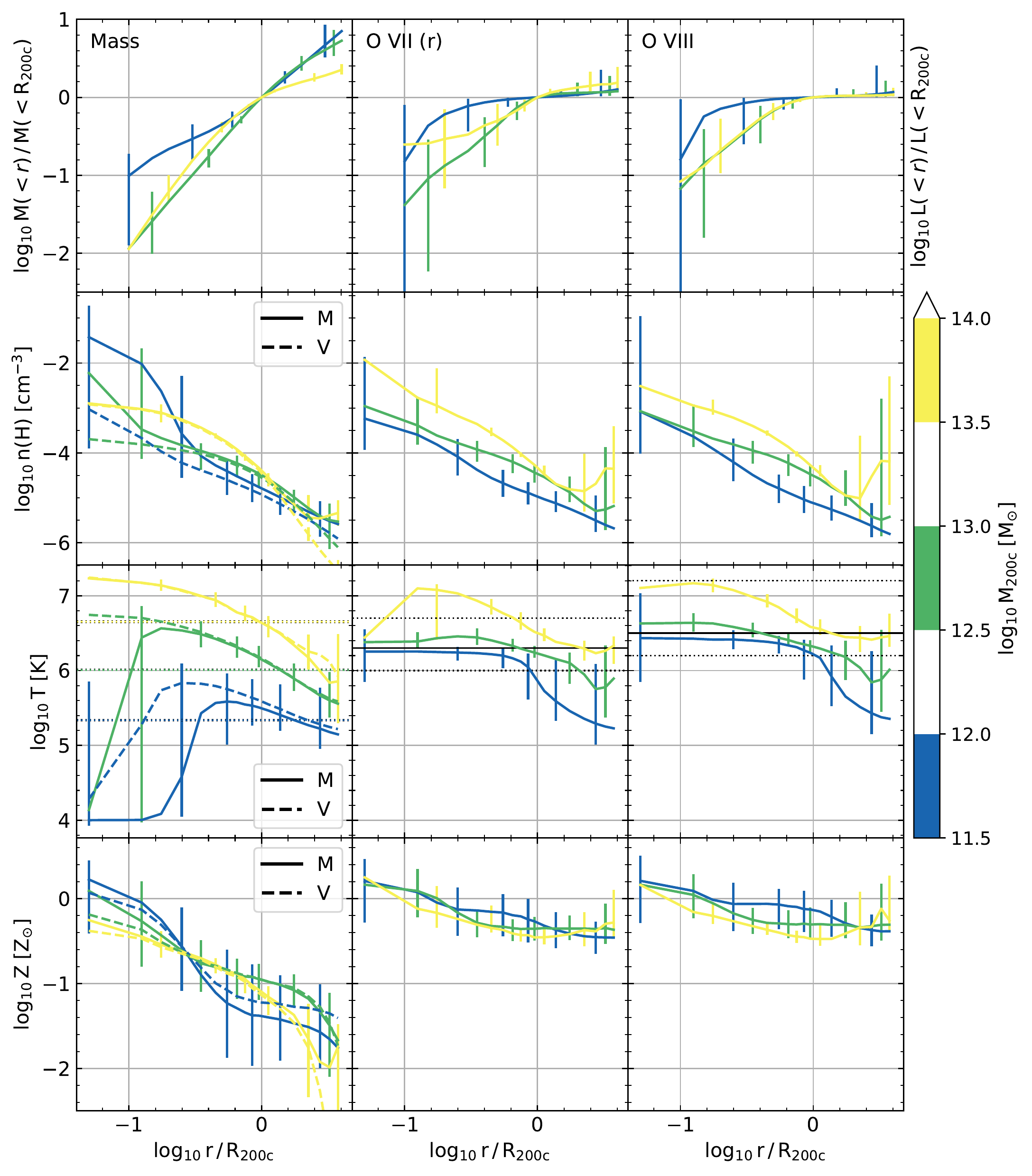}
\caption{Radial profiles of enclosed gas mass and line luminosity (top row), and mass-, volume-, and luminosity-weighted gas temperature (second row), density (third row), and metallicity (fourth row). Mass- and volume-weighted median quantities are shown with solid and dashed lines, respectively. The columns correspond to mass/volume (left),  \ion{O}{vii}~(r) (middle) and \ion{O}{viii} (right). The median profiles are the medians of the individual halo median profiles. The error bars indicate the 10$^{\txn{th}}$ and 90$^{\txn{th}}$ percentiles of the individual halo medians, i.e., the inter-halo scatter. For the cumulative emission, the percentiles were taken after normalizing by the enclosed luminosity (or mass) within $\Rvir$.}
\label{fig:3dprof_em}
\end{figure*}

To investigate which gas in haloes is responsible for the line emission, we examine emission-weighted temperatures, densities, and metallicities as a function of distance to the central galaxy. For each emission line, we first make a histogram of the SPH particles around each individual central galaxy, binning them by distance to the central galaxy (normalized by $\Rvir$) and e.g., temperature, weighted by luminosity. 
To combine these into a profile for the whole halo mass bin, we first extract the luminosity-weighted median temperatures in each radial bin for individual haloes. For the cumulative emission profiles, we extract the cumulative emission profiles for individual haloes and normalize each individual profile to the enclosed luminosity within $\Rvir$. We then extract the median of these medians at each radius and in each halo mass bin.
Fig.~\ref{fig:3dprof_em} shows these medians of medians at each halo mass and radius. We show the median temperature, hydrogen number density, and metallicity (oxygen mass fraction) in each radial bin, for each halo mass bin. The enclosed luminosities 
$\mathrm{L(<\mathrm{r})} \,/ \, \mathrm{L}_{\mathrm{200c}}$ are the medians of the normalized individual halo enclosed luminosties. The error bars indicate the 10$^{\txn{th}}$ and 90$^{\txn{th}}$ percentiles, in alternating radial bins for legibility. We do not show the scatter for the volume-weighted medians, but it is similar to that of the mass-weighted medians. 



\begin{figure*}
\includegraphics[width=0.8\textwidth]{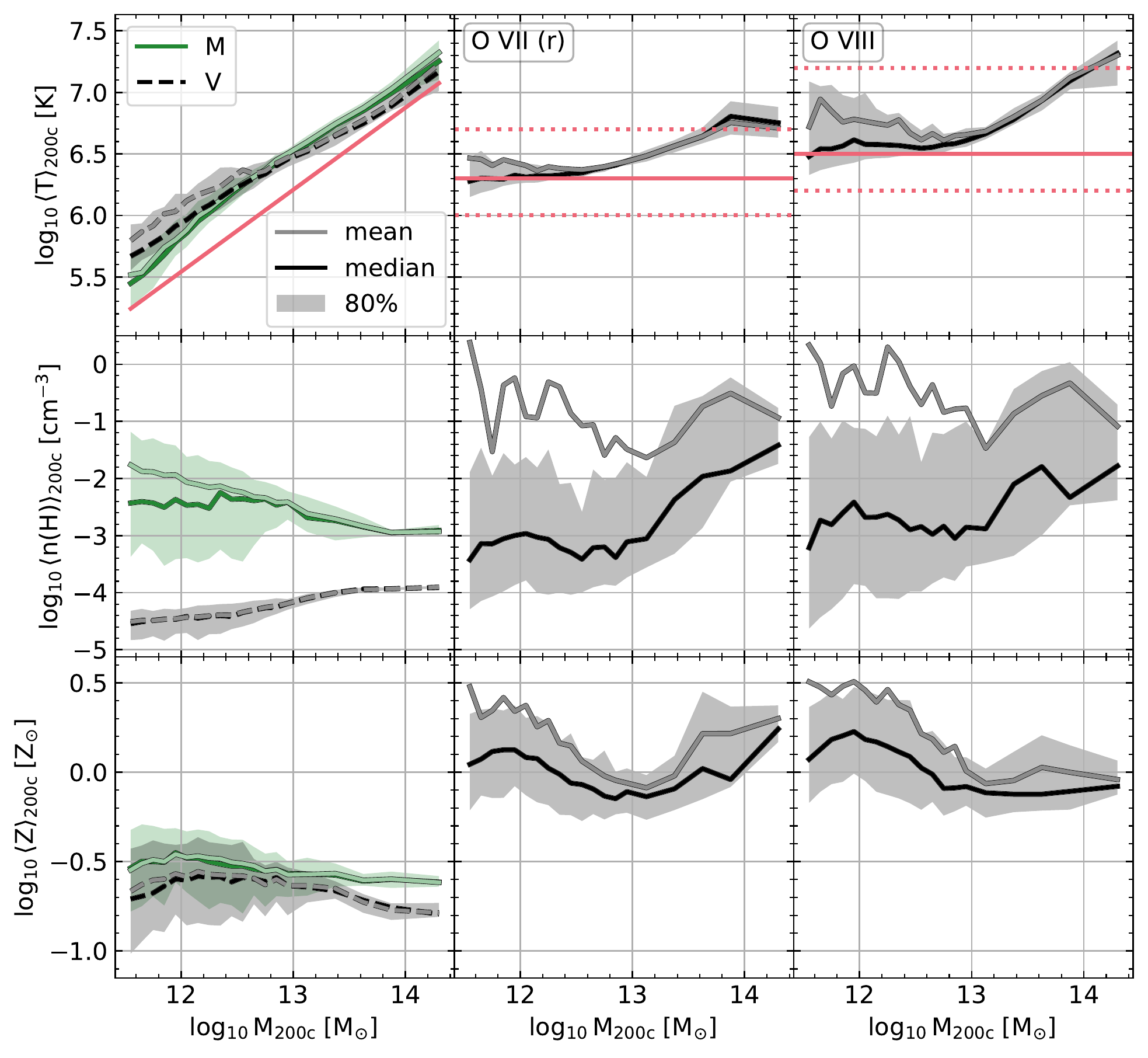}
\caption{Mass-, volume-, and luminosity-weighted gas temperature (top row), density (middle row), and metallicity (bottom row) for all non-star-forming gas within $\Rvir$ as a function of halo mass. Dark lines and shading show the median and scatter (80~per~cent) of the individual halo averages, respectively, while lighter lines show weighted averages over the individual halo values at each halo mass. From left to right, different columns show quantities weighted by mass/volume , \ion{O}{vii}~(r) luminosity and \ion{O}{viii} luminosity. Mass- and volume-weighted quantities are shown in solid, green and black, dashed lines, respectively. Red lines in the temperature plots (top row) show $\Tvir$ in the left panel, while in the remaining panels of the top row they indicate the peak CIE temperature (solid) and the range where the emissivity is at least 0.1 times the maximum in CIE (dashed).}
\label{fig:Lw}
\end{figure*}

The middle and right columns of Fig.~\ref{fig:3dprof_em} show the emission-weighted profiles for \ion{O}{vii}~(r) (middle column) and \ion{O}{viii} (right column). These can be compared to the similarly obtained mass- and volume-weighted gas properties shown in the left column.
These profiles include all gas; star-forming gas is included at $10^{4}$~K. 
The data in the first column of Fig.~\ref{fig:3dprof_em} are the similar to the first column of \citet[][figs.~12 and~13]{wijers_schaye_oppenheimer_2020}\footnote{The individual halo properties are combined in a different way to obtain the profiles of \citet{wijers_schaye_oppenheimer_2020}, but the results for the mass- and volume-weighted profiles are similar.}.

In the top row of Fig.~\ref{fig:3dprof_em}, we see the enclosed mass and luminosity as a function of distance to the central galaxy in different halo mass bins. We see that within $\Rvir$, the emission tends to be more concentrated in halo centres than the mass. 

The emission-weighted metallicities are similar to those found by \citet{van-de-voort_schaye_2013} using the OWLS simulations. The trends are also similar to those for the parent ion-weighted metallicities of \citet{wijers_schaye_oppenheimer_2020}; indeed, emission and absorption share the same $\propto Z$ scaling, so a similar bias would be expected. Emission is biased towards high-metallicity gas. The emission-weighted median metallicity is higher than the mass- or volume-weighted value, and declines less strongly with radius. The difference with the mass- and volume-weighted metallicities reflects the substantial scatter in gas metallicity at large radii. Note that for the gas metallicity, the displayed inter-halo scatter is smaller than the scatter within haloes (median of individual halo 10$^{\txn{th}}$ and 90$^{\txn{th}}$ percentiles). Depending on the emission line, the emission-weighted metallicity varies between declining and mostly flat with radius.  
Note that for the density and temperature profiles, the inter- and intra-halo scatter are generally similar.

The emission-weighted densities tend to broadly follow the trend of the volume-weighted profiles, but show a bias towards higher densities. Again, this is consistent with the findings of \citet{van-de-voort_schaye_2013}. The bias is particularly large at high halo masses and outside $\Rvir$, where the mass- and volume-weighted densities are lowest. Such a bias is expected given that emission scales with the density squared. However, we do notice that where the mass- and volume-weighted densities differ, at small radii for lower-mass haloes, the emission-weighted densities are lower than the mass-weighted densities. This is because those high densities coincide with too low temperatures (the cool ($\sim 10^{4}$~K), dense gas phase) to produce significant emission in these soft X-ray lines (Fig.~\ref{fig:emcurves}).

Indeed, the emission-weighted temperatures are consistently high, and are not very sensitive to the overall gas temperature. Rather, this emission traces whatever gas is present around its emissivity peak temperature. 
However, we do see some emission from below the emissivity peak where gas densities are low around the lowest-mass haloes we consider, suggesting photo-ionization is a factor there. We note that most of the emission within $\Rvir$ comes from radii where collisional processes dominate. This also agrees with the findings of \citet{van-de-voort_schaye_2013}.


In Fig.~\ref{fig:Lw}, we explore these trends as a function of halo mass. Here, the temperature, density, and metallicity are averaged within $\Rvir$ for each halo, and we show the trends of these weighted averages with halo mass. As Fig.~\ref{fig:3dprof_em} would suggest, the mass- and volume-weighted gas temperatures roughly follow $\Tvir$, but are somewhat larger. 

For the emission lines, we find trends for He-$\alpha$-like resonance lines and K-$\alpha$ lines that are illustrated by the \ion{O}{vii} and \ion{O}{viii} lines we show. The Fe L-shell lines follow trends similar to the He-$\alpha$-like lines. 
The emission-weighted temperature lies close to the CIE emissivity peak.
The emission-weighted temperature does follow the halo virial temperature trend over a limited mass range, but within the constraints of the line emissivity peak. 


The volume-weighted density reflects the halo (non-star-forming) gas fraction.
We see that this increases with halo mass. Interestingly, the scatter decreases around the same halo mass where the scatter in line luminosity decreases (Fig.~\ref{fig:Ltot}). This is in line with the trend \citet{davies_crain_etal_2019} found with total soft X-ray luminosity (their fig.~4), where the scatter at fixed halo mass was driven by the halo gas fraction, especially for haloes hosting galaxies with stellar luminosity $\gtrsim \Lstar$. The mass-weighted density remains high relative to the volume-filling density even at the largest halo masses.

We see a generally rising trend of emission-weighted median density with halo mass, with emission tracing higher densities than mass at high halo masses, where the virial temperature exceeds the emissivity peak temperature. For the different lines, the emission-weighted density becomes roughly constant with halo mass in the regime where the emission-weighted temperature is similar to the emissivity peak temperature. 
For the K-$\alpha$ lines, and some other lines with relatively high peak emissivity temperatures, the emission-weighted mean temperature falls above the 80~per~cent halo-to-halo scatter range here, which is typically true for the density in a much larger halo mass range. This indicates that the brightest haloes here differ considerably from the typical haloes in the gas that causes their emission. Comparing radial profiles obtained by combining individual halo data in different ways\footnote{These profiles were obtained by directly adding the emission-weighted radius-temperature histograms of individual haloes and taking the emission-weighted median at each radius, and by doing the same, but normalizing each halo's histogram by the luminosity enclosed within $\Rvir$ before adding them.} (not shown), suggests that this difference is driven by emission-weighted temperatures in the halo centres. The emission in the brightest haloes at these low masses is often driven by direct heating of gas by feedback, meaning the luminosity predictions in these brightest, low-mass haloes are less reliable (Appendix~\ref{app:directfb}).

The mass-weighted metallicities are likely higher than the volume-weighted ones simply because dense gas tends to be closer to the galaxies where the metals are produced. 
The emission-weighted average halo metallicities are of order $\txn{Z}_{\odot}$, which is well above the mass- and volume-weighted ones; this is expected, since metal-line emission is inevitably biased towards metals. 

For the lines with narrower emissivity peaks (He-$\alpha$-like and Fe L-shell), the emission-weighted metallicity tends to increase with halo mass starting roughly where the halo gas temperature (mass or volume weighted) exceeds the peak emissivity temperature of the line. For the K-$\alpha$ lines, the metallicity tends to flatten out at this mass instead. At similar halo masses, the emission-weighted densities rise. Hence, the larger the factor by which the typical temperature exceeds the value for which the emissivity peaks, the more highly biased the emission-weighted density and metallicity tend to become. 



\section{Discussion}
\label{sec:discussion}

\subsection{The EAGLE simulations}

Current hydrodynamical simulations lack the resolution to model feedback processes from first principles and must hence make use of subgrid models that are calibrated to some observables. In the case of large-volume simulations such as {\eagle}, the model is calibrated to the observed low-redshift galaxy mass function and sizes. However, models with widely varying in- and outflow rates can result in the same galaxy masses \citep[e.g.,][]{mitchell_schaye_etal_2020, mitchell_schaye_2022}. Indeed, CGM predictions can vary dramatically between simulations that reproduce the galaxy mass function. 

For example, \citet{davies_crain_etal_2019_tngcomp} compared the gas mass content of the CGM in {\eagle} and IllustrisTNG~100-1. They found that at halo masses $M_{200} \gg 10^{12} \Msun$, the {\eagle} CGM contains a somewhat higher gas fraction, while the IllustrisTNG CGM contains much more gas at masses $\lesssim 10^{12} \Msun$. In fact, the IllustrisTNG gas fractions have a minimum (as a function of halo mass) at $M_{200} \sim 10^{12.5} \Msun$. In {\eagle}, the gas fraction consistently increases with halo mass, although the slope does change at $M_{200} \sim 10^{12.5} \Msun$. This difference in low-mass halo gas fraction, and corresponding soft X-ray luminosity \citep{davies_crain_etal_2019}, likely drives differences in \ion{O}{vii} emission from $\Mvir \approx 10^{11.5} \Msun$ haloes found by \citet{simionescu_ettori_etal_2019}. 

\citet{eagle_paper} found that the gas fractions in high-mass {\eagle} halos ($\mathrm{M}_{500, \mathrm{hse}} > 10^{13.5} \Msun$) are too high at fixed halo masses. The soft X-ray luminosity is $\approx 0.3 \dex$ too high for fixed spectroscopic temperatures $\gtrsim 1$~keV. \citet{barnes_kay_etal_2017} investigated the X-ray properties in more detail and found that these most massive haloes in the EAGLE $100^{3} \cMpc^{3}$ Reference model contain too much gas at fixed $\mathrm{M}_{\mathrm{500c}}$, and are a bit too cool. The soft X-ray luminosities (0.5--2~keV) are about right though, as are the metallicities (iron), so the metal emission line predictions might be realistic despite the simulation's flaws. For halo masses $\lesssim 10^{13} \Msun$, X-ray observations are currently insufficiently sensitive to test the predictions. 

The sensitivity of the CGM to in- and outflows \citep[e.g.,][]{mitchell_schaye_2022} makes it a useful testbed for models of galaxy formation, which motivates studies like ours. \citet{oppenheimer_bogdan_etal_2020} predict that, with eRosita stacking, the difference between the IllustrisTNG~100-1 and {\eagle} CGM soft X-ray emission predictions for nearby $\sim \Lstar$ galaxies should be detectable, as well as the connection between quenching and halo gas fraction (via the central galaxy star formation rate and soft X-ray surface brightness).  

If the numerical resolution is changed in a large-volume galaxy formation simulation like {\eagle}, the subgrid prescription effectively changes since it moves to a different scale and will generally result in different CGM gas flows \citep[see the discussion in \S2 of ][]{eagle_paper}. Hence, we expect the predictions for CGM emission to also change with the resolution of the simulation. This will remain true even if the subgrid parameters are recalibrated to match the galaxy mass function, since we know that calibration on galaxy properties leaves room for a wide range of CGM predictions.

We test the effect of simulation resolution on the surface brightness profiles shown in Fig.~\ref{fig:medmeanprof} in Appendix~\ref{app:resconv}. 
We test for this using a recalibrated, higher-resolution version of the {\eagle} simulation, run in a $25^{3} \cMpc^{3}$ volume: \code{Recal-L025N0752} \citep{eagle_paper}. This simulation has 8 (2) times better mass (spatial) resolution than our fiducial simulation \code{Ref-L100N1504}. Because we are testing the resolution dependence in a smaller volume, our sample of high-mass haloes is very small. There are no haloes with $\Mvir > 10^{13.5} \Msun$, and only one with $\Mvir > 10^{13} \Msun$.

For haloes with $\Mvir \approx 10^{11.5}$--$10^{13} \Msun$, the properties of the CGM depend somewhat on the resolution and its implications for feedback, but these effects are relatively small. For those haloes, the median and mean surface brightness profiles typically differ by $< 0.5 \dex$ between the simulations with these different resolutions, across the different emission lines. This difference is small compared to the range of surface brightness values in the 0.1--1~$\Rvir$ impact parameter range. The high-resolution median surface brightnesses tend to be higher than the \code{Ref-L100N1504} values (those in Fig.~\ref{fig:medmeanprof}), meaning the predictions in Fig.~\ref{fig:medmeanprof} for the detectability of individual haloes are, in this sense, conservative. 
 
 At lower halo masses ($\Mvir \lesssim 10^{11.5} \Msun$), the intrinsic properties of the haloes (CGM gas fraction and temperature) differ more between haloes at different resolutions, and the convergence of the mean surface brightness profiles becomes poorer, particularly in the central regions. This motivates the range of halo masses we show throughout this work.

\subsection{Literature comparison}
Other predictions of CGM soft X-ray emission lines have been made. 
\citet{van-de-voort_schaye_2013} used the $100 h^{-1} \cMpc^{3}$ OWLS simulations \citep{schaye_dalla-vecchia_etal_2010} to predict the CGM emission from a number of soft X-ray emission lines (\ion{C}{IV}, \ion{N}{VII}, \ion{O}{VII}, \ion{O}{VIII}, and \ion{Ne}{X}), and compared these to estimated detection limits of a set of X-ray instruments. 
They used different halo mass bins and have more high-mass haloes due to their larger simulation volume. We note that their mass resolution is nearly two orders of magnitude lower than for EAGLE and that their fiducial model does not include AGN feedback.

Although these differences make direct comparison difficult, a few trends are clearly similar. The hierarchy of line brightnesses for the five soft X-ray lines is similar, and the brightnesses fall in a similar range. The shapes of the profiles are, however, different. The \citet{van-de-voort_schaye_2013} profiles show a central core at $\Mvir = 10^{12}$--$10^{13} \Msun$, while the surface brightness continues to rise towards the smallest radii at  $\Mvir = 10^{13}$--$10^{14} \Msun$, and there is a central peak in surface brightness at $\Mvir = 10^{14}$--$10^{15} \Msun$. We see a trend in the opposite direction: the lowest-mass haloes have the most centrally peaked emission, while at $\Mvir = 10^{13}$--$10^{14} \Msun$, there is more of a core within $\sim 0.1 \Rvir$. Though our $\Mvir > 10^{14} \Msun$ sample is small (9 haloes), we see a clear dip in surface brightness in the centres of these most massive haloes. This is physically plausible because the centres of these {\eagle} haloes are their hottest parts (Fig.~\ref{fig:3dprof_em}), and these haloes are hotter than ideal for producing these lines overall (Fig.~\ref{fig:emcurves}). We saw a similar effect in soft X-ray absorption in \citet{wijers_schaye_oppenheimer_2020}.

\citet{simionescu_ettori_etal_2019} compare predictions for \ion{O}{vii} CGM emission in IllustrisTNG~100-1 \citep{pillepich_springel_etal_2018} and {\eagle} in their fig.~3. The {\eagle} profiles were calculated with a different set of line emission tables than we use. This should, however, not make a big difference for the predictions, because the emission mostly comes from CIE gas, so the UV/X-ray background is not important\footnote{Median profiles for the K-$\alpha$ and He-$\alpha$-like lines computed from the two sets of tables we use in this paper differ by $\approx 0.1 \dex$ in the potentially observable (surface brightness $> 10^{-2} \, \mathrm{photons} \, \mathrm{s}^{-1} \mathrm{cm}^{-2} \mathrm{sr}^{-2}$) regime. This is consistent with the differences we find between the emissivities as a function of temperature in CIE.}. 
The {\eagle} and IllustrisTNG predictions are similar at a halo mass of $10^{12.5} \Msun$, but differ substantially at $10^{11.5} \Msun$: the IllustrisTNG haloes are much brighter in their centres, but the emission drops off more rapidly with impact parameter, leaving the {\eagle} haloes brighter at the virial radius. Note that these low-mass {\eagle} haloes are not detectable in \ion{O}{vii} emission at any radius,
at least with the instruments considered in this work. The predictions from the Illustris simulation \citep[the predecessor of  IllustrisTNG;][]{vogelsberger_genel_etal_2014_methods} 
differ substantially from the {\eagle} and  IllustrisTNG predictions at both halo masses.

In agreement with \citet{van-de-voort_schaye_2013},
we find that metal emission-line-weighted metallicities are biased high relative to mass- and volume-weighted metallicities, across the halo masses we consider (Fig.~\ref{fig:3dprof_em}). The bias increases with distance from the central galaxy, as the line-weighted values are a roughly constant $Z \approx 0.3$--$1 \us \mathrm{Z}_{\odot}$ outside $\approx 0.3 \Rvir$, while the mass- and volume-weighted metallicities decrease with distance to the central galaxy out to at least $\approx 3 \Rvir$, reaching $Z \approx 0.03$--$0.1 \us \mathrm{Z}_{\odot}$ at $\Rvir$. 
These emission-line-weighted metallicities are similar to the metallicities 
\citet{barnes_kay_etal_2017} found from mock, (broadband) X-ray observations of their C-EAGLE clusters ($\mathrm{M}_{\mathrm{500c}} = 10^{13.9}$--$10^{15.1} \Msun$). These are a set of simulated clusters, which use a variation of the {\eagle} code similar to the \code{Reference} model we used in this work: \code{AGNdT9} \citep[][]{eagle_paper}. The values they find from their mock observations are consistent with metallicities measured from observations.

Various metallicity measurements from ICM emission spectra indeed indicate that the metallicity (iron) of the ICM is roughly constant from $\approx 0.3 \Rvir$ to the largest radii where there are measurements, 
$\sim \Rvir$ \citep[e.g., fig.~3 of the review by][]{mernier_biffi_etal_2018}.
\citet{martizzi_hahn_etal_2016} studied cluster (halo mass $\sim 10^{15} \Msun$) metallicities using a different set of simulations. Their X-ray-emissivity-weighted metallicities were lower than metallicities measured from observations, but the metallicity bias is similar to what we find in lower-mass haloes in {\eagle}: it increases with distance to the central galaxy. The emissivity-weighted values are roughly constant, while the mass-weighted metallicity decreases with distance to the central galaxy. They attribute this difference to the fact that their metals are concentrated in dense gas clumps at large distances. (Their emissivity weighting is based on bremsstrahlung density and temperature scalings, and does not depend on metallicity.)
\citet{biffi_planelles_etal_2018} similarly found flatter metallicity profiles when weighting by emission instead of mass; they additionally used 3-dimensional distances for their mass-weighted profiles and impact parameters for the emission-weighted profiles. Their emission-weighted profiles matched observations.

In the CGM of simulated $\sim \Lstar$ galaxies, \citet{crain_mccarthy_etal_2013} found a similar metallicity bias in broadband X-ray emission. Like what we find for X-ray emission lines, this bias increases with distance to the central galaxy, as luminosity-weighted metallicities remain constant around $\Rvir$, or decrease less strongly with distance than the mass-weighted values.

Our results suggest that the biases in metallicity measurements from ICM X-ray emission extend to the CGM of haloes that are three orders of magnitude less massive than those clusters. This highlights the value of numerical simulations in the interpretation of observational findings. We do note that we compare mass- and volume-weighted metallicities to  values weighted by metal line emission, not total X-ray emission. This likely results in at least somewhat larger differences (biases) than would result from observations.    

We have found that there is considerable scatter in the luminosities and surface brightnesses of haloes at fixed halo mass (Figs.~\ref{fig:Ltot} and~\ref{fig:medmeanprof}). The work of \citet{davies_crain_etal_2019} describes a likely driver of this scatter. They examined the scatter in the total soft X-ray (0.5--2~keV) luminosity  of haloes in the {\eagle} simulation. The main driver of this scatter was found to be the amount of energy AGN had injected into the gas (measured through the black hole mass). Haloes that had experienced more feedback were left with lower halo gas fractions, and therefore less and lower density gas to produce X-ray emission. The AGN feedback also quenches star formation, resulting in a positive correlation between star formation rate and X-ray luminosity at fixed halo mass. The soft X-ray emission will be dominated by emission lines at halo masses below the cluster range \citep[e.g., the review by][]{werner_mernier_2020}. Therefore, it is reasonable to assume that the emission lines will be affected by AGN feedback in a similar way to the total soft X-ray emission. This would result in an anti-correlation of emission-line luminosity with black hole mass at fixed halo mass, and a positive correlation with star formation rate at fixed halo mass.

In previous papers, we examined X-ray absorption lines using the {\eagle} simulations \citep{wijers_schaye_etal_2019, wijers_schaye_oppenheimer_2020}. Compared to X-ray line absorption, line emission has a much stronger density bias. This means that while line absorption traces both collisionally and photo-ionised gas \citep[][figs.~11--13]{wijers_schaye_etal_2019}, line emission almost exclusively traces collisionally ionised gas (Fig.~\ref{fig:pds}). This difference in density bias also means that line emission is much more concentrated in haloes than the ions producing the lines \citep[compare Fig.~\ref{fig:Lsplit} to fig.~2 of][]{wijers_schaye_oppenheimer_2020}. While absorption lines can be used to detect both halo gas and IGM \citep{wijers_schaye_oppenheimer_2020}, emission lines cannot be used to detect gas outside haloes in single structures. We note that we have not investigated stacking or statistical detection methods aimed at IGM gas.

\citet{rahmati_etal_2016} had similarly investigated a number of UV absorption lines, including the \ion{O}
{vi} and \ion{Ne}{viii} lines probing the cooler part of the warm-hot gas. Their fig.~7 is comparable to our Fig.~\ref{fig:pds} and fig.~13 of \citet{wijers_schaye_etal_2019}. These show that the X-ray lines generally probe warmer (in CIE) and lower-density (in PIE) gas than the UV lines. However, there is considerable overlap in the gas probed by the \ion{Ne}{viii} and \ion{O}{vii} ions. These UV ions are less concentrated in haloes than the X-ray ions and line emission, and both lines are detectable over a larger range of halo masses than the X-ray absorption and emission lines \citep{wijers_schaye_oppenheimer_2020}.

\section{Conclusions}
\label{sec:conclusions}

We have investigated soft X-ray metal-line emission from the CGM and the IGM in the $100^{3} \cMpc^{3}$ {\eagle} simulation, for a sample of bright lines including the brightest ones we expect. We investigated K-$\alpha$ and He-$\alpha$-like emission lines, and a few iron L-shell lines, with rest-frame energies between 0.3 and 2~keV and emissivity peaks in CIE between $\approx 10^6$ and $10^7$~K. Our main conclusions about the line emission are:

\begin{itemize}
\item{Line emission is dominated by haloes, i.e., CGM, rather than by the interhalo IGM (Figs.~\ref{fig:igmstamps} and \ref{fig:Lsplit}). The emission is more concentrated in haloes than the ions producing this emission, where we have data for both \citep[\ion{O}{vii}, \ion{O}{viii}, \ion{Ne}{ix}, and \ion{Fe}{xvii} from][]{wijers_schaye_oppenheimer_2020}. The difference is most likely due to the stronger density dependence of emission compared to ion density.}
\item{The brightest emission comes from the \ion{O}{viii} K~$\alpha$ doublet, and the other K~$\alpha$ lines have bright peak surface brightnesses as well. The brightest He-$\alpha$-like resonance lines come from \ion{O}{vii}. The Fe L-shell lines reach peak surface brightnesses similar to or somewhat larger than that of the \ion{O}{vii} resonance line, in a narrower range of halo masses (Fig.~\ref{fig:medmeanprof}).}
\item{There is large scatter in line luminosity at fixed halo mass. The scatter decreases towards higher halo masses and median luminosities (Fig.~\ref{fig:Ltot}).}
\item{Line emission originates mainly from gas at CIE temperatures, even far from the central galaxy. For K-$\alpha$ lines, emission can originate from hotter gas in high-mass haloes (where $\Tvir$ is above the emissivity peak temperature). Photo-ionization by the UV/X-ray background may be important in some low-mass haloes, but it does not seem to matter for emission from haloes we might be able to detect in line emission (Fig.~\ref{fig:Lw}).} 
\item{Line emission is biased to high-metallicity gas (Figs.~\ref{fig:3dprof_em} and~\ref{fig:Lw}), in agreement with the findings of \citet{van-de-voort_schaye_2013}. This bias is similar to what we found for metal absorbers in \citet{wijers_schaye_oppenheimer_2020}.
Others have found similar metallicity biases for broadband X-ray emission in clusters \citep[e.g.,][]{martizzi_hahn_etal_2016, barnes_kay_etal_2017, biffi_planelles_etal_2018} and the CGM of $\sim \Lstar$ galaxies \citep{crain_mccarthy_etal_2013}.}
\item{We have also examined trends of the halo luminosity and surface brightness in various emission lines with halo mass. The primary driver of these trends is how close the temperature of the halo ($\sim \Tvir$) is to the temperature where the emissivity of the line peaks. This is the `virial temperature thermometer' effect discussed by \citet{oppenheimer_etal_2016} in the context of \ion{O}{vi} absorption.}
\item{Secondarily, the shape of the emissivity curve (as a function of temperature in CIE) matters. For the Fe L-shell lines, the emissivity peaks are narrow (Fig.~\ref{fig:emcurves}), and surface brightnesses depend strongly on halo mass. For the He-$\alpha$-like lines, the emissivity peaks are less narrow, and the dependence of surface brightness on halo mass is less strong. The K-$\alpha$ lines have the widest peaks, with emissivity decreasing slowly towards high temperatures. The surface brightnesses of these lines generally keep increasing with halo mass, and only start to peak or plateau at $\sim 10^{14} \Msun$, where the sample size is severely limited by the volume of the simulation.}
\end{itemize}

We also assessed the prospects for detecting line emission from the CGM with different instruments. We did this by calculating simplified and generally optimistic estimates of minimum observable surface brightnesses (\S\ref{sec:det}). We ignore any systematic errors and define emission as detectable if it would constitute a $5 \sigma$ detection as determined from the signal to noise ratio. For the noise, we include estimates of instrumental and astrophysical backgrounds. We use the limits for exposure times and spatial binning $\Delta t \, \Delta \Omega = 1$ and $10$~Ms~arcmin$^{2}$. We compare these detection limits to the surface brightness profiles of typical CGM emission and stacked CGM emission in Fig.~\ref{fig:medmeanprof}. 

\begin{itemize}
\item{With the XRISM Resolve instrument, we will likely be able to observe some CGM emission in the brightest lines, from haloes with $\Mvir \gtrsim 10^{13}$--$10^{13.5} \Msun$.}
\item{With the Athena X-IFU and the Lynx Main Array, it will be possible to detect line emission from haloes down to $\Mvir \approx 10^{12}$--$10^{12.5} \Msun$. For haloes with $\Mvir \gtrsim 10^{13} \Msun$ it may even be possible to detect the outer CGM of the haloes in \ion{O}{viii} and \ion{O}{vii} emission lines, with very long exposure times (1--10~Ms) or stacking. The inner CGM of $\sim \Lstar$ galaxies may also be accessible with these two ions, long exposures, and stacking. }
\item{For emission lines below $\approx 1$~keV, the Lynx Ultra-High Resolution Array will provide increased sensitivity. With this instrument, imaging the inner CGM of galaxies down to $\sim \Lstar$ masses will be possible, without stacking but with long exposure times, in \ion{N}{vii} and \ion{C}{vi} K-$\alpha$ emission. In \ion{O}{viii} and \ion{O}{vii} line emission, less extreme exposure times or spatial binning are expected to be sufficient.}
\end{itemize}

\section*{Acknowledgements}

We thank Luigi Piro for help with the X-IFU detection limits, Aurora Simionescu for help with the XRISM Resolve limits, and Alexey Vikhlinin for help with the Lynx limits. We thank Ben Oppenheimer for useful discussions.

We used the \textsc{python} \textsc{sherpa} package \citep{freeman_doe_siemiginowska_2001, doe_nguyen_etal_2007} and \textsc{astropy}\footnote{\url{http://www.astropy.org}} \citep{astropy_2013, astropy_2018} to handle the response matrices and backgrounds for detection limits. Other \textsc{python} packages we used include \textsc{numpy} \citep{numpy_2020}, \textsc{scipy} \citep{scipy_2020}, \textsc{h5py} \citep{h5py}, and \textsc{matplotlib} \citep{matplotlib}, and the \textsc{ipython} \citep{ipython} command-line interface.
We thank Paul Tol for making his colour blind friendly colour schemes publicly available.

This work is partly funded by Vici grant 639.043.409 and research programme Athena 184.034.002 from the Dutch Research Council (NWO).
This paper is supported by the European Union's Horizon 2020 research and innovation programme under grant agreement No 871158, project AHEAD2020.
This work used the DiRAC@Durham facility managed by the Institute for Computational Cosmology on behalf of the STFC DiRAC HPC Facility (www.dirac.ac.uk). The equipment was funded by BEIS capital funding via STFC capital grants ST/K00042X/1, ST/P002293/1, ST/R002371/1 and ST/S002502/1, Durham University and STFC operations grant ST/R000832/1. DiRAC is part of the National e-Infrastructure.

\section*{Data availability}

The {\eagle} halo and galaxy catalogues \citep{mcalpine_helly_etal_2016} and the complete simulation outputs \citep[snapshots;][]{eagle-team_2017} are publicly available at \url{http://icc.dur.ac.uk/Eagle/database.php}. The response files and backgrounds for the different instruments are available as indicated in the text. Data and plots from this work are available from the corresponding author on reasonable request.





\bibliographystyle{mnras}
\bibliography{bibliography} 



\appendix

\section{Gas directly heated by feedback}
\label{app:directfb}

\begin{figure*}
\includegraphics[width=\textwidth]{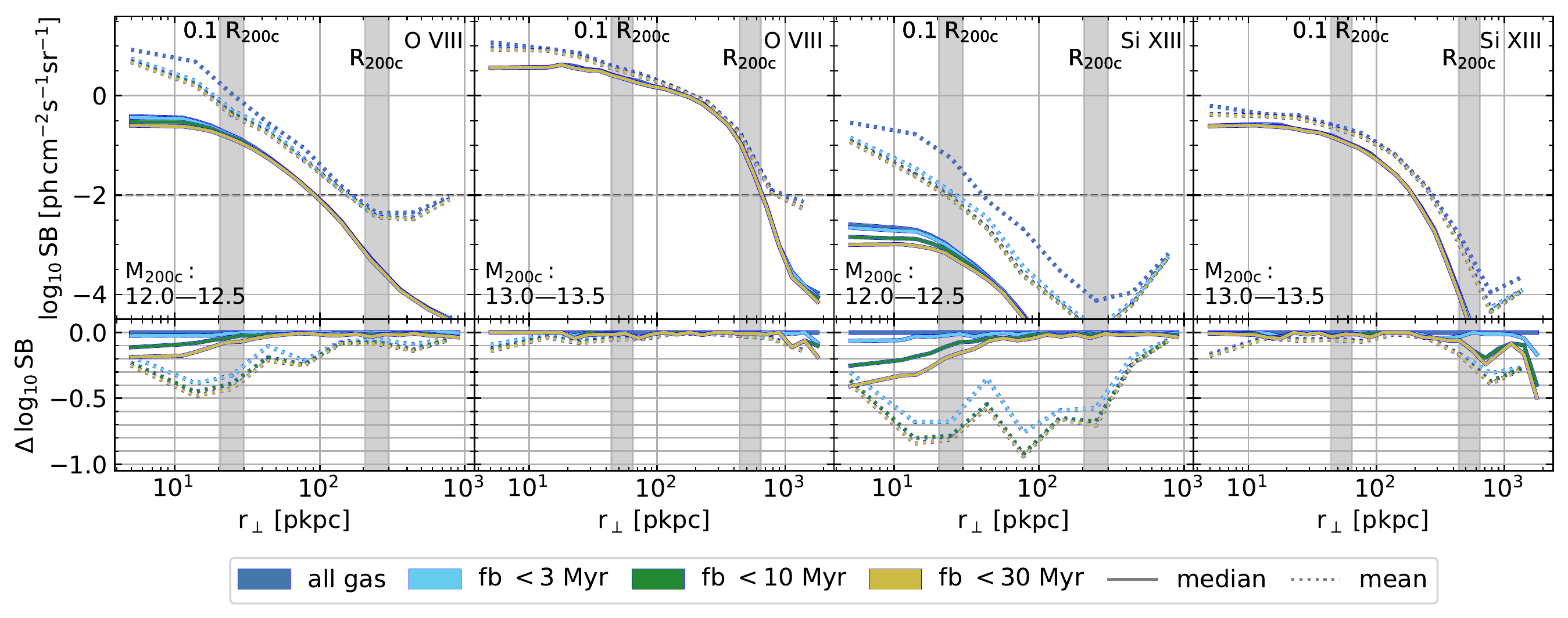}
\caption{Examples of the effect that gas particles that have recently been directly injected with feedback energy have on emission line surface brightness profiles. The dark blue lines (`all gas') match those of Fig.~\ref{fig:medmeanprof}. Solid lines are medians of the annular average profiles around individual haloes, dotted lines show the means of these profiles. The lines in other colours show the profiles obtained by excluding gas directly heated by feedback, less than 3, 10, or 30~Myr ago. 
We show the profiles for two emission lines: \ion{O}{viii} in the left two panels and \ion{Si}{xiii} in the right two panels. For each emission line, we show two halo mass ranges: 
$\Mvir = 10^{12}$--$10^{12.5} \Msun$ in the leftmost and centre-right panels, and $\Mvir = 10^{13}$--$10^{13.5} \Msun$ in the centre-left and rightmost panels. The emission lines and halo mass ranges (in $\log_{10} \Msun$) are indicated in the panels.
The top panels show the surface brightness profiles, the bottom panels show the differences with the respective mean and median `all gas' profiles. The directly heated gas can have a substantial effect on the emission in the halo centre ($\lesssim 0.1 \Rvir$), and at halo masses where the virial temperature is too low for the emission line to be strong. While the effect on the medians is modest, it can be large for the mean profiles. At halo masses sufficiently high for the virial temperature to be $\gtrsim$ the peak emissivity temperature, the effect of directly heated gas is small.}
\label{fig:directfb}
\end{figure*}

In this appendix, we investigate the effect of gas that has been directly heated by stellar or AGN feedback on the surface brightness profiles.  Feedback in {\eagle} is implemented by a stochastic energy injection, causing a fixed temperature increase of $10^{7.5}$ or $10^{8.5}$~K in the directly heated gas particles for supernovae and AGN, respectively. These values are motivated by numerical considerations and calibration of galaxy population properties, not by expected physical temperatures of e.g., supernova bubbles, which remain unresolved. Therefore, if the surface brightness profiles we find were dominated by this directly heated gas, then the profiles may not be a realistic prediction of the {\eagle} simulation.

We test the effect of this directly heated gas by making profiles excluding it. For this, we use the maximum past temperature of each gas particle, and the redshift at which that maximum was achieved, which are tracked by the simulation. We refine our selection by inspecting phase diagrams: the distribution of gas mass in density-temperature space. We compare all gas in the simulation to the distribution of gas that has maximum temperatures 
$\log_{10} \, \mathrm{T}$ -- $\log_{10} \, \mathrm{T} + \Delta$
corresponding to each type of feedback as a function of the time since the maximum temperature was reached. Much more gas is directly heated by supernovae than by AGN, and its temperatures are closer to the emissivity peak temperatures of our emission lines, so the details of the AGN-heated gas selection are less important than those of the supernova-heated gas. Using the phase diagrams, we estimate which maximum temperatures and time lags include the high-density and high-temperature gas that has just been heated, and not too much of the gas that forms the bulk of the mass distribution in {\eagle}. This is because, after enough time has passed, the predictions for the temperature of the gas reflect the properties of the bulk outflows and are less sensitive to the precise manner in which the energy was injected into individual particles. 
We assume that gas that reached a maximum temperature between $10^{7.5}$ and $10^{7.7}$~K was heated by supernova feedback,  and that a maximum between $10^{8.5}$ and $10^{8.7}$~K means the gas was heated by AGN feedback. However, the supernova feedback temperatures can also be reached by virialized gas at high halo masses (Fig.~\ref{fig:3dprof_em}). We estimate that gas at densities $\mathrm{n}_{\mathrm{H}} \lesssim 10^{-2} \pcc$, and temperatures $\gtrsim 10^{7.4}$~K, is part of a continuous distribution of gas, heated by e.g., virial shocks instead of supernovae. Therefore, we do not exclude this diffuse gas from the `no direct heating' profiles. 

We show the resulting profiles, excluding gas that was heated less than 3, 10, or 30~Myr ago, in Fig.~\ref{fig:directfb}. To illustrate the general trends, we show profiles for two emission lines and two halo masses. Firstly, in the halo centres (impact parameters $\lesssim 0.1 \Rvir$) the effects of the directly heated gas can be quite large for both the mean and median profiles. However, this is the region where, in observations, the emission from the CGM would be difficult to distinguish from that of the central galaxy (e.g., the hot ISM). Secondly, although at larger impact parameters (up to $\Rvir$) the effects of direct heating can be severe for mean profiles, this is limited to halo masses which produce little emission overall in that emission line. 

We also looked at other lines and halo masses than plotted in Fig.~\ref{fig:directfb}. For impact parameters $\approx 0.1$--$1 \Rvir$, we find that the differences in the mean and median profiles are typically not worse than those in the leftmost panel of Fig.~\ref{fig:directfb} at all halo masses we investigate for the carbon, nitrogen, and oxygen lines. 
For the neon and iron L-shell lines, and the \ion{Mg}{xi}~(r) line, this difference threshold is met in $\Mvir \gtrsim 10^{12.5} \Msun$ haloes. For the \ion{Mg}{xii} K-$\alpha$ and \ion{Si}{xiii}~(r) lines, the threshold lies at $\Mvir \gtrsim 10^{13} \Msun$. 

This means that, at halo masses for which we predict the CGM to be observable (median profiles in Fig.~\ref{fig:medmeanprof}), our predictions are not very sensitive to the direct heating of gas by feedback. Where haloes only seem to be observable within $0.1 \Rvir$ (typically marginally), the surface brightnesses might however be artificially high due to the way feedback is implemented in {\eagle}. The same is true for halo masses that seem observable only in mean stacks, especially in halo centres, but where the stacked mean surface brightness is much higher than the median surface brightness.



\section{Numerical convergence}
\label{app:resconv}

In this section, we discuss the convergence of the surface brightness profiles with the resolution of the simulation.
In order to test this, we compare surface brightness profiles from two {\eagle} volumes:
\code{Ref-L025N376}, and \code{Recal-L025N0752}
\citep{eagle_paper}.
Both have a volume of $25^{3} \cMpc^{3}$, which is smaller than the $100^{3} \cMpc^{3}$ of the main simulation we use throughout this work 
(\code{Ref-L100N1504}).
 The mass (spatial) resolution of the
 \code{Recal-L025N0752} simulation is 8 (2) $\times$ higher than that of \code{Ref-L100N1504}. Its feedback parameters were calibrated in the same way as those of \code{Ref-L100N1504}, but at its higher resolution. The \code{Ref-L025N0376} uses the same resolution and feedback prescription as \code{Ref-L100N1504}, but in the same volume and using the same initial conditions as
\code{Recal-L025N0752}.

The comparison between the \code{Ref-L025N0376} and \code{Recal-L025N0752} models tests the `weak convergence' of the emission profiles, in the terminology of \citet{eagle_paper}. This is based on the idea that, even at fixed parameters, the effect of subgrid feedback will typically depend on the scale at which it is injected, and therefore on the resolution of the simulation. In that context, a resolution test cannot be seen independently of the feedback model, and a simulation using a similar calibration at higher resolution provides a fair test of resolution convergence.

\begin{figure*}
    \centering
    \includegraphics[width=\textwidth]{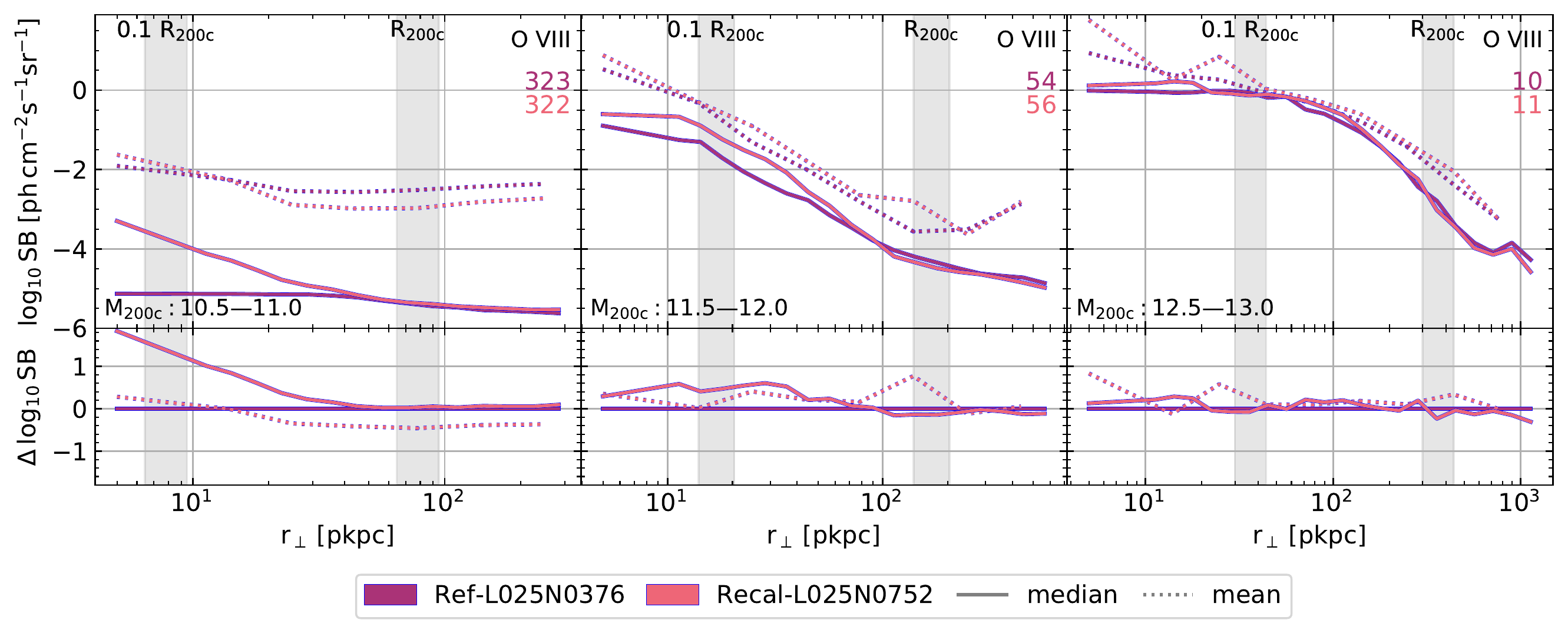}
    \caption{A comparison of the surface brightness profiles from the \code{Ref-L025N0376} and the \code{Recal-L025N0752} simulations.
    The \code{Ref-L025N0376} simulation uses the same model and resolution as the \code{Ref-L100N1504} simulation used throughout this work, but it has a volume of $25^{3} \cMpc^{3}$ instead of $100^{3} \cMpc^{3}$.
    \code{Recal-L025N0752} is an {\eagle} simulation with its feedback parameters recalibrated at its $8 \times$ higher mass resolution, and the same $25^{3} \cMpc^{3}$ volume as \code{Ref-L025N0376}. The number of objects in each $\Mvir$ bin is shown at the top right of each panel. We show profiles for \ion{O}{viii} in three halo mass bins. The mass ranges are indicated in the bottom left of the panels, in $\log_{10} \Msun$. These \ion{O}{viii} profiles are representative of the level of convergence for other emission lines. Except for the mean profile at small radii and low halo masses, the results are reasonably converged.}
    \label{fig:resconv}
\end{figure*}

We illustrate the level of convergence in Fig.~\ref{fig:resconv}, where we compare the profiles for the \ion{O}{viii} K$\alpha$ doublet as an example, which is representative of the level of convergence at a given halo mass for other potentially observable emission lines.
In short, the profiles are well-converged in haloes of mass $\Mvir \approx 10^{12.5}$--$10^{13} \Msun$. In haloes with $\Mvir \lesssim 10^{11.5} \Msun$ convergence is however poor for the mean profiles in the central regions. In these low-mass haloes the CGM has not developed a hot, virialized phase \citep[e.g.,][]{dekel_birnboim_2006, keres_katz_etal_2009, van-de-voort_schaye_etal_2011, correa_schaye_etal_2018}, leading to very low surface brightnesses, below the predicted detection limits. For haloes with $\Mvir \approx 10^{11.5}$--$10^{12.5} \Msun$, convergence is reasonable given the range of surface brightnesses within $\Rvir$. Differences
of $\approx 0.5 \dex$ remain, but these are small compared to the decline in surface brightness with radius and compared with the differences between the mean and median profiles. 


Across halo masses and emission lines, the \code{Recal-L025N0752} median surface brightness predictions tend to be higher than the \code{Ref-L025N0376} values. In this sense, the Fig.~\ref{fig:medmeanprof} predictions for detectability of individual haloes in soft X-ray line surface brightness are conservative. 

The halo temperature, density, and metallicity, and their emission-line-weighted values as shown in Fig.~\ref{fig:Lw} are reasonably converged at $\Mvir \gtrsim 10^{11.5} \Msun$. Some small differences remain: the \code{Recal-L025N0752} haloes are typically slightly cooler and contain slightly more gas, especially at $\Mvir \lesssim 10^{12.5} \Msun$.


\bsp	
\label{lastpage}
\end{document}